\documentclass[12pt]{article}
\usepackage{a4}
\usepackage{amsfonts}
\usepackage{amssymb}
\usepackage{cite}
\usepackage{epsfig}
\usepackage[all]{xy}
\usepackage{amsmath}
\usepackage{cite}
\author{}

\newcommand{\be}{\begin{equation}}
\newcommand{\ee}{\end{equation}}
\newcommand{\ba}{\begin{array}}
\newcommand{\ea}{\end{array}}
\newcommand{\bea}{\begin{eqnarray}}
\newcommand{\eea}{\end{eqnarray}}

\newcommand{\ov}{\overline}
\def\IR{\relax{\rm I\kern-.18em R}}

\def\IP{\relax{\rm I\kern-.18em P}}
\def\inbar{\vrule height1.5ex width.4pt depth0pt}
\def\IC{\relax\,\hbox{$\inbar\kern-.3em{\rm C}$}}

\newcommand{\Z}{\mathbb{Z}}

\def\K3{{\bf K3}}

\def\ov{\overline}

\def\n2d{\cN_{V^*}^{\otimes 2}}

\def\IC{\mathbb{C}}

\def\IR{\mathbb{R}}

\def\IP{\mathbb{P}}

\def\cN{{\mathcal N}}

\begin{document}

\title{
\begin{flushright} \vspace{-2cm}
{\small UPR-1194-T\\
} \end{flushright}
\vspace{4.0cm}
(Non-)BPS bound states and D-brane instantons}
\vspace{1.0cm}
\author{\small  Mirjam Cveti{\v c}, Robert Richter, Timo Weigand}

\date{}

\maketitle

\begin{center}
\emph{Department of Physics and Astronomy, University of Pennsylvania, \\
     Philadelphia, PA 19104-6396, USA }
\vspace{0.2cm}

\tt{cvetic@cvetic.hep.upenn.edu, rrichter@physics.upenn.edu, timo@physics.upenn.edu}
\vspace{1.0cm}
\end{center}
\vspace{0.7cm}

\begin{abstract}

We study non-perturbative effects  in four-dimensional ${\cal N}=1$ supersymmetric orientifold compactifications due to  D-brane instantons which are not invariant under the orientifold projection.
We show that they can yield superpotential contributions via a multi-instanton process at threshold. Some constituents of this configuration form bound states away from the wall of marginal stability which can decay in other regions of moduli space.
A microscopic analysis reveals how contributions to the superpotential are possible when new BPS states compensate for their decay.
We study this concretely for D2-brane instantons along  decaying special Lagrangians in Type IIA and for D5-branes instantons carrying holomorphic bundles in Type I theory.

\end{abstract}

\thispagestyle{empty}
\clearpage


\section{Introduction}

Quantum corrections to the superpotential of four-dimensional ${\cal N}=1$ supersymmetric string vacua
are interesting in theory and practice.
Due to non-renormalisation of the superpotential at the perturbative level, non-perturbative effects play, though exponentially suppressed, a crucial role in that they can represent the leading-order contributions of certain couplings in the effective action.
The revived recent interest, starting with \cite{Blumenhagen:2006xt,Haack:2006cy,Ibanez:2006da,Florea:2006si}, in D-brane instanton effects  has its origin precisely in this fact. So-called stringy or exotic D-brane instantons wrapping cycles not necessarily populated by a spacetime-filling brane can yield  various types of perturbatively absent couplings in the effective action of phenomenological significance \cite{Blumenhagen:2006xt,Haack:2006cy,Ibanez:2006da,Florea:2006si,Abel:2006yk,Akerblom:2006hx,Bianchi:2007fx,Cvetic:2007ku,Argurio:2007qk,Argurio:2007vq,Bianchi:2007wy,Ibanez:2007rs,Akerblom:2007uc,Antusch:2007jd,Blumenhagen:2007zk,Aharony:2007pr,Blumenhagen:2007bn,Aharony:2007db,Billo:2007sw,Billo:2007py,Aganagic:2007py,Camara:2007dy,Cvetic:2007qj,Ibanez:2007tu,GarciaEtxebarria:2007zv,Petersson:2007sc,Blumenhagen:2007sm,Bianchi:2007rb,Matsuo:2008nu,Blumenhagen:2008ji,Argurio:2008jm}.
These are to be contrasted with conventional gauge instantons, whose realisation as D-brane instantons was investigated in detail in \cite{Billo:2002hm,Billo:2006jm}.
Ground-breaking early work on D-brane instantons appeared in \cite{Becker:1995kb,Witten:1995gx,Douglas:1995bn,Green:1997tv} including the computation of certain multi-instanton effects in setups with extended supersymmetry \cite{Ooguri:1996me,Bachas:1997mc,Kiritsis:1997hf,Bachas:1997xn}.

Consider for definiteness an ${\cal N}=1$ supersymmetric orientifold compactification of Type II string theory.
The common lore is that for a Euclidean D-brane to correct the superpotential it has to wrap a suitable BPS cycle of
the compactification manifold whose BPS phase is aligned with that of the orientifold plane.
The BPS condition guarantees that the instanton is a volume minimizing representative of its homology class. In this sense it constitutes a local minimum of the string action, thereby fulfilling the analogue of the defining characteristic for gauge instantons.

In backgrounds where ${\cal N}=1$ as opposed to ${\cal N}=2$
supersymmetry is preserved locally along the internal space, such as
M-theory on G2 manifolds or heterotic compactifications, Euclidean
half-BPS objects break two of the four supercharges of the effective
field theory, and the associated Goldstone fermions $\theta^{\alpha}$ enable the
object to generate F-terms \cite{Witten:1996bn}. In Type II
orientifolds, due the local enhancement of supersymmetry from ${\cal
N}=1$ to ${\cal N}=2$  away from the orientifold plane, BPS
instantons generically carry four such universal Goldstinos,
$\theta^{\alpha}$ and $\ov \tau^{\dot \alpha}$
\cite{Argurio:2007qk,Argurio:2007vq,Bianchi:2007wy,Ibanez:2007rs}.
For suitable instantons invariant under the orientifold action the
anti-chiral ones are projected out, and the way is paved, in
principle, for the generation of a superpotential.
Instantons invariant under the orientifold projection in such a way that only two Goldstinos $\theta^{\alpha}$ survive are called $O(1)$ instantons.
Another important class of Euclidean D-branes is given by gauge instantons, which wrap the same cycle as a
spacetime filling brane. Here the extra Goldstinos $\ov \tau^{\dot \alpha}$ are needed to implement the
ADHM constraints \cite{Billo:2002hm}. This can be generalised to
instantons along a single spacetime filling D-brane even though the associated $U(1)$
gauge group has no field theoretic gauge instantons
\cite{Aganagic:2007py,Petersson:2007sc}. In the rest of this article we will be concerned with stringy instantons in the sense that
they wrap a cycle not populated by any other D-brane of the compactification.

The above stated conditions for superpotential contributions  prompt two immediate questions. First, what is the role of the much more generic BPS instantons in orientifold models which are not invariant under the orientifold action, called $U(1)$ instantons in the sequel? Second, are BPS cycles the only source for corrections to the superpotential, or do non-BPS instantons contribute as well?

In \cite{Blumenhagen:2007bn}, an analysis of the first question was initiated. $U(1)$-instantons are best described in the upstairs geometry where they are given by a pair of instantons wrapping the cycle $\Xi$ and its orientifold image $\Xi'$.
As we will review in section \ref{sec_def}, the original reason to discard such $U(1)$-instantons, the two extra anti-chiral Goldstone modes $\ov \tau^{\dot \alpha}$, does not necessarily withstand closer scrutiny. In favourable circumstances, the coupling of the  $\ov \tau^{\dot \alpha}$ modes to other instanton zero modes in the instanton effective actions allows for their absorption in the instanton path integral, leaving us with two zero modes $\theta^{\alpha}$ in the universal sector. If the instanton really contributes to the superpotential depends on the absence of extra other unliftable modes. For a rigid cycle, such extra zero modes can arise in the sector between the instanton and its orientifold image or between the instanton and the D-branes of the compactification. One of the results of \cite{Blumenhagen:2007bn} is that for $U(1)$ instantons of chiral intersection
type with its orientifold image, global constraints always enforce the presence of charged zero modes of the latter type.

This is unfortunate as it is precisely this chiral sector that is
related also to the second question regarding the role of non-BPS
instantons. As we will review in some detail
 in section \ref{sec_boundstate1}, the BPS condition for cycles is known to depend on the closed string moduli. BPS cycles can become marginally stable along lines of marginal stability in moduli space and disappear upon passing this hypersurface. The above cycles wrapped by $U(1)$ instantons of a chiral type are precisely of that form. Their study in the region of moduli space where they exist as properly calibrated BPS cycles can thus give us some insights into the role of non-BPS cycles. The reason is that by holomorphicity, the instanton induced superpotential has to be of the same functional form on both sides of the line of marginal stability. This was discussed in \cite{GarciaEtxebarria:2007zv} for the special case of a line of threshold stability where BPS cycles become marginally stable with respect to its constituents without necessarily decaying across the hypersurface. Rather, another BPS state of the same charges forms on the other side which can now contribute to the superpotential. $U(1)$ instantons with non-minimal intersection with its image are of this type.

Continuity of quantum corrections across lines of marginal stability despite jumps in the responsible BPS spectrum is a known phenomenon in gauge and string theories with ${\cal N}=2$ supersymmetry (see e.g.\cite{Sethi:1996em,Sethi:1997pa,Dorey:2001ym,Dorey:2002ik}). The closest analogue of instanton generated superpotential terms in Type II orientifolds is given by instanton corrections to the hypermultiplet metric in the parent Type II compactifications \cite{Halmagyi:2007wi}.

In the present paper, with this motivation in mind, we revisit possible superpotential contributions of a $U(1)$ instanton $\Xi$ with chiral intersection with its image $\widetilde \Xi$. We show that, while due to the presence of extra charged zero modes no single instanton contributions are possible, these modes can be lifted in a multi-instanton process involving another two $O(1)$ instantons $\widetilde \Xi_1$ and $\widetilde \Xi_2$. Perturbing this system away from the line of marginal stability for the $U(1)$-instanton and its image, on one side a multi-instanton involving a BPS bound state between  $\Xi$ and  $\Xi'$ takes over in generating a superpotential. On the other side, by contrast, this BPS object does not exist. However, the additional instantons  $\widetilde \Xi_1$ and $\widetilde \Xi_2$ conspire to form a different BPS bound state of the same total charge that contributes to the superpotential.

We also discuss a slightly simpler multi-instanton configuration where at threshold all extra fermionic zero modes are lifted, but no BPS object exists upon deforming the moduli. This is consistent as even at threshold
a superpotential contribution is impossible by the vanishing integral over the bosonic moduli space.
The way in which this non-BPS cycle violates the BPS condition is somewhat subtle. Based on its associated effective field theory, we argue that it is destabilised by linear F-term obstructions involving massive adjoint fields in the open string sector. Before turning to the instanton analysis, in section \ref{sec_F-term_obst} we describe this mechanism in general as we find it interesting in itself. Along the way we propose that D-brane instantons can lead to a quantum deformation of the BPS spectrum of a compactification.

The detailed discussion of $U(1)$ instantons, their bound states and decay in section \ref{MultiIIA} is given in the language of Euclidean $D2$-instantons of general Type IIA Calabi-Yau orientifolds.  To make sure we are not working on the empty set, we construct an example of the configuration we have in mind on $T^6/{\mathbb Z}_2 \times {\mathbb Z}'_2$. Due to its technical character we relegate its presentation to Appendix \ref{App_IIA}.

Our results suggest that the class of instantons correcting the superpotential is larger than commonly appreciated. To further demonstrate this, we translate the IIA setup of section \ref{MultiIIA} into Type I compactifications in section \ref{sec_TypeI}. Here it is Euclidean $D5$-branes carrying certain vector bundles that become relevant in addition to the usually studied $D1$-instanton corrections. The formation of bound states of our multi-instanton configuration can be described quite explicitly in the language of extensions. Finally we also reconsider the problem of vector-like intersections of a $U(1)$ instanton and its image and analyse under what conditions extra zero modes in the $E-E'$ sector are lifted.
Some more technical details can be found in Appendix \ref{app_Lifting}.
Section 5 contains our conclusions.

\section{Decay of BPS states across lines of marginal stability}

\subsection{Bound state decay in absence of F-term obstructions}
\label{sec_boundstate1}

We begin by recalling some basic facts about (bound states of) BPS branes and their decay which we will make frequent use of in this article. For a review and the standard references on this vast and fascinating subject see e.g. \cite{Aspinwall:2004jr}.

Compactify Type IIA or IIB string theory on a Calabi-Yau threefold
$X$. Supersymmetric D-branes are given by topological A- or B-type
branes, respectively, which are stable in a suitable sense. Their
notion is encoded in the concept of the Fukaya category and the
derived bounded category of coherent sheaves \cite{Douglas:2000gi},
respectively. In the geometric phase, the relevant A-type branes are
given by Lagrangian three-cycles\footnote{More generally,
coisotropic branes in the sense of  \cite{Kapustin:2003se} (see also
\cite{Font:2006na}).}, while at large volume B-type branes can be
thought of as holomorphic cycles carrying holomorphic bundles or
sheaf theoretic generalisations thereof. Note that the definition of
topological A- and B-type branes involves the K\"ahler and complex
structure moduli, respectively.

The BPS condition on the other hand comes in two parts.
In order to preserve an ${\cal N}=1$ subalgebra of the ${\cal N}=2$ supersymmetry preserved by the Calabi-Yau, topological branes have to satisfy a stability criterion. For A-type branes, this is the special Lagrangian condition \cite{Becker:1995kb}, while in the B-type case the sheaves have to be stable with respect to a suitably defined slope \cite{Marino:1999af,Douglas:2000ah,Bridgeland:2005my}.
Associated with such BPS objects is a central charge $Z$, which in the large volume regime reads
\bea
\label{equ_Z}
 Z \simeq \left\{\begin{array}{cc} 
\int_{\Pi} \Omega  & {\rm A-type \,\, branes} \\
\int_X e^J {\rm ch}(i {\cal F}) \sqrt{{\rm td}(X)} & {\rm B-type \, \,branes} \, \end{array} \right\}.
\eea
Note that $Z$ depends only on the complex structure $\Omega$ or the K\"ahler structure $J$ for A- or B-type branes, respectively. The quoted expression for B-type branes refers to branes wrapping the whole of $X$ and carrying a bundle with curvature ${\cal F}$.
The particular ${\cal N}=1$ supersymmetry preserved by the BPS brane is parameterised by the phase
\bea
\varphi = {\rm Arg}(Z).
\eea
In a configuration with several D-branes, a common ${\cal N}=1$ supersymmetry is only preserved once all BPS phases are aligned.
In Calabi-Yau orientifolds the orientifold plane singles out a preferred ${\cal N}=1$ subalgebra, and
in what follows we will set the associated reference phase to $0$.

The equation fixing the phase of BPS branes in agreement with the
orientifold plane is related to the D-flatness conditions of the
four-dimensional effective field theory supported by
spacetime-filling branes. For small deviations from a supersymmetric
configuration, $|\varphi| <<1$, the breaking of ${\cal N}=1$
supersymmetry by non-aligned BPS states can be described as
spontaneous D-term breaking within the usual two-derivative
supergravity framework. Otherwise, higher derivative terms become
non-negligible. In the above limit one can identify the BPS phase
with the Fayet-Iliopoulos term of the diagonal $U(1)$ subgroup
associated with the D-brane theory, \bea 2 \pi \alpha' \xi =
\varphi. \eea

Consider for simplicity the abelian low-energy effective theory of a pair of BPS D-branes $\Pi_1$ and $\Pi_2$  with $n^+$ and $n^-$ chiral fields of positive and negative charge with respect to $U(1)_1 - U(1)_2$ \cite{Kachru:1999vj}.
Both BPS branes preserve the same ${\cal N}=1$ supersymmetry provided the D-term
\bea
\label{D-term0}
V_D = \frac{1}{2 g_{YM}^2} \left( \sum_i |q _i| |\phi^+_i|^2 - \sum_j |q _j| |\phi^-_j|^2  - \xi \right)^2
\eea
vanishes.
For zero vacuum expectation values (VEVs) of the charged scalar fields, the supersymmetry condition $\xi =0$ singles out a real codimension 1 hypersurface in complex or K\"ahler moduli space which we will denote by ${\cal M}_0$ in the sequel.
On this locus, there exists a BPS object with homological charges $[\Pi_1] + [\Pi_2]$, given by $\Pi_1 \cup \Pi_2$.

Deforming the respective moduli away from ${\cal M}_0$ generates an FI term $\xi$, and according to its sign we enter into the regions of moduli space denoted by ${\cal M}_-$ or ${\cal M}_+$.
In  ${\cal M}_+$ the fields $\phi^+_i$, if present, are tachyonic and their condensation  can trigger the formation of a bound state which we denote by $\Pi_2 \# \Pi_1$. The existence of this bound state is guaranteed only in a small neighbourhood away from ${\cal M}_0$. Likewise, in ${\cal M}_-$, condensation of $\phi^-_i$ can lead to formation of the BPS bound state $\Pi_1 \# \Pi_2$. The charge of each of these bound states is again  $[\Pi_1] + [\Pi_2]$. In the limit of sufficiently small deformations away from ${\cal M}_0$, the FI terms (or BPS phases) of the constituent objects add up linearly upon bound state formation.

We have to distinguish the following qualitatively different cases: If $n^+ = n^- \neq 0$, i.e. for vector-like intersections, BPS bound states exist on either side of ${\cal M}_0$, which should therefore be called, adopting the nomenclature of \cite{Denef:2001ix,deBoer:2008fk}, line of threshold stability.
The same is true for chiral intersections where $0 \neq n^+ \neq n^- \neq 0$. By contrast, the interesting case of strictly chiral intersections with either $n^+ \neq 0$ or $n^- \neq 0$ leads to the genuine decay of a BPS object, say the bound state $\Pi_2 \# \Pi_1$ in ${\cal M}_+$, as we pass the line of marginal stability, where  the $\Pi_1 \cup \Pi_2$ is BPS.
In general the representatives of a given homological charge can meet several lines of marginal and/or threshold stability in moduli space.

For special Lagrangians, bound states are described geometrically by the connect sum of their constituents \cite{Joyce:1999tz}, while for B-type branes bound state formation is encapsulated in the distinguished triangles of the derived category \cite{Aspinwall:2004jr}. For our purposes it is enough to think of bound states as a non-split extension. For early work in this context see \cite{Sharpe:1998zu,Thomas:2001ve}.  We will adopt this viewpoint in section \ref{sec_TypeI}.

\subsection{(Non-)BPS bound states and  F-term obstructions}
\label{sec_F-term_obst}
{\bf 1.) Classical example} \\

In more general situations, F-terms can destabilise otherwise BPS objects or obstruct the formation of BPS bound states.
As an illustration consider the following simple
system of 3 single spacetime-filling BPS-branes $D_a$, $D_b$, $D_c$
which are taken to be suitable BPS A- or B-type branes, respectively. The associated field theory was considered before in \cite{Maillard:2007pq, Kumar:2007dw} as a model of supersymmetry breaking.
If all three branes are calibrated with respect to the orientifold, the low-energy effective field theory  is ${\cal N}=1$ SYM with gauge group $U(1)_a \times U(1)_b \times U(1)_c$ (modulo one decoupled overall $U(1)$). We assume that the charged matter content of the system is given just by three chiral superfields $\Phi_{(-1_a,1_b)}$, $B_{(-1_b,1_c)}$ and $A_{(-1_c,1_a)}$.

Starting from the situation where all three branes preserve the same ${\cal N}=1$ supersymmetry as the orientifold, we are interested in the behaviour of the BPS-branes upon infinitesimal deformations of the complex or K\"ahler structure, respectively. We are considering only such deformations for which the brane $D_c$ continues to preserve the same ${\cal N}=1$ supersymmetry as the orientifold, i.e.
the FI-term associated with $U(1)_c$ vanishes, $\xi_c =0$.

For sufficiently small deformations, the behaviour of the system is captured by the scalar potential of the effective field theory,
\bea
V=V_D + V_F,
\eea
where
\bea
V_D &\simeq& \frac{1}{2g_{YM}^2} \left( ( - |\phi|^2 + |A|^2 - \xi_a) ^2 + (  |\phi|^2 -|B|^2 - \xi_b) ^2 + (|B|^2- |A|^2)^2 \right), \nonumber \\
V_F &\simeq&  \lambda^2 \, (|\Phi|^2 |B|^2 + |\Phi|^2 | A|^2 +  |A|^2 |B|^2).
\eea
Here we consider, for simplicity, equal gauge couplings for all 3 branes. $\lambda$ denotes the Yukawa coupling appearing in the superpotential $W= \lambda \, \Phi B A$, which we assume to be non-vanishing.

Unbroken SUSY is possible only for $-\xi_b =  \xi_a = \xi \leq 0$, and the microscopic behaviour of the branes in this regime is clear.
Perturbing the system instead such that  $-\xi_b =  \xi_a = \xi > 0$, we have the following non-SUSY minimum for perturbatively small values of $x={2g_{YM}^2} \, {\lambda^2}$:
\bea
|A| = |B| = \sqrt{\frac{2\xi}{2+x}}, \quad\quad \Phi=0.
\eea
F- and D-flatness are both broken as the F-term prevents the system from recombining into a D-flat configuration corresponding to
$|A| = |B| = \sqrt{\xi}$.

To understand this, we first consider  the hypothetical BPS-bound state $\Psi$ due to condensation of the tachyons $A$  and $B$ in absence of the F-term.
It can be viewed as the result of first condensing $A$, leading to the intermediate state
$Y = D_c \# D_a$, and its subsequent combination with $D_b$ induced by the VEV of $B$,
\bea
\Psi = D_b \# D_c \# D_a.
\eea

Due to the described linearity of the FI-terms in the limit of small deformations, the  would-be BPS bound state $\Psi$ leads to a vanishing D-term, in agreement with the field theory analysis for $|A| = |B| = \sqrt{\xi}$.
It still hosts a massless chiral multiplet $\Phi$ playing the role of a modulus, while the adjoint fields $A$, $B$ have acquired D-term masses. But the F-term $W = \lambda \, \Phi   B A $ before bound state formation indicates that the modulus $\Phi$ is actually 'obstructed' at linear order in that it suffers from a tadpole
$W = \lambda \xi \Phi$. Together with the coupling $\lambda \sqrt{\xi} \, \Phi \, (\delta A + \delta B)$ to the massive fluctuations $\delta A$, $\delta B$ this tadpole  leads, in the scalar potential, to destabilising terms linear in $\delta A$, $\delta B$. The bound state $\Psi$ is driven into a truly non-BPS  state $\widetilde \Psi$ of the same homological charge which breaks both D- and F-flatness while minimizing the total action.

Geometrically, it is not completely obvious in which sense  $\widetilde \Psi$ violates the BPS condition. We would like to argue that it is not just a calibrated cycle preserving the wrong ${\cal N}=1$ subalgebra, but rather not calibrated at all.
After all, for calibrated cycles the BPS-phase depends only on the charges, see equ. (\ref{equ_Z}).
So  we cannot form another BPS bound state in the same homology class as $\Psi$ but with a different BPS phase.
On the other hand, we see no indications that $\widetilde \Psi$ ceases to satisfy the topological brane, i.e. Lagrangian or holomorphicity, condition.
Its violation should manifest itself in extra closed moduli dependent F-terms in the effective action (see e.g. \cite{Martucci:2006ij}) in addition to the matter potential.
We therefore propose that $\widetilde \Psi$ is a \emph{non-calibrated A- or B-type brane}, respectively.
The presence of the destabilising superpotential terms for the hypothetical cycle $\Psi$ reflects the fact that the geometry does actually not allow for a stable BPS cycle of this charge in this region of moduli space.
\\

{\bf 2.) D-instanton generated F-term obstructions} \\

The above situation is an example of a 'classical'  obstruction of a BPS brane in that the responsible F-terms arise at string tree-level. More generally such F-terms can be induced by  stringy effects due to D-brane instantons.
Consider e.g. a system of two BPS branes $D_a$, $D_b$ with bifundamental matter $\Phi$ and a corresponding D-term
\bea
V_D = \frac{1}{2g_{YM}^2} \left( |\Phi|^2 - \xi \right)^2.
\eea
Much like in the example before, the formation of a BPS bound state $ \Psi = D_b \# D_a$ for  $\xi > 0$
can be obstructed e.g. by a
quadratic F-term of the form
\bea
\label{quad_term}
W = m \, \Phi^2.
\eea
Such superpotential terms are generated by stringy D-brane instantons wrapping suitable BPS cycles which intersect the D-branes \cite{Blumenhagen:2006xt,Ibanez:2006da,Florea:2006si}.
In \cite{Aharony:2007db} this mechanism was considered as a realisation of the Fayet model of spontaneous supersymmetry breaking.
Our point of view here is that the D-brane instanton responsible for (\ref{quad_term}) leads to a \emph{quantum deformation of the geometry} in the sense that it induces a linear obstruction in the scalar potential for the massive adjoint $\Phi$ of the would-be BPS bound state $D_b \# D_a$. In the same spirit as in the above classical example the BPS state $\Psi$ is destabilised towards formation of a non-calibrated brane and thus disappears from the quantum corrected BPS spectrum.

This quantum deformation of the BPS spectrum depends in an interesting way on the global properties of the string compactification and not merely on the local details of the geometry. The point is that the instanton inducing (\ref{quad_term}) might intersect in addition some other D-branes. In this case there are extra charged fermionic zero modes between the instanton and these other D-branes. They have to be absorbed by bringing down from the instanton action their couplings, if present, to other modes $\widetilde \Phi_i$ in the D-brane sector which do not arise at the intersection $D_a-D_b$. The coupling (\ref{quad_term}) is modified to
\bea
W' \simeq \Phi^2 \, \times \prod_i \widetilde \Phi_i
\eea
and need not destabilise the BPS bound state  $D_b \# D_a$ (provided the operator $\prod_i \widetilde \Phi_i$ does not take a non-zero VEV in the vacuum). It would be interesting to study this effect further.

\section{Chiral instanton recombination as a multi-instanton process}
\label{MultiIIA}
\subsection{Definition of setup}
\label{sec_def}

After this preparation we finally turn to the analysis of
superpotential contributions of four-dimensional ${\cal N}=1$
Calabi-Yau orientifold compactifications
\cite{Blumenhagen:2005mu,Blumenhagen:2006ci} from so-called $U(1)$
instantons, as defined in the introduction. These were first studied
systematically in \cite{Blumenhagen:2007bn}. We discuss a
prototypical configuration in the context of a Type IIA
compactification on a Calabi-Yau $X$ modded out by the combined
action $\Omega \ov \sigma$ of worldsheet parity $\Omega$ and an
anti-holomorphic involution $\ov \sigma$ acting on $X$. The mirror
symmetric Type IIB picture will be described in section
\ref{sec_TypeI}.

Let $E$ and $E'$ denote a Euclidean $D2$-brane\footnote{These are
dubbed $E2$-instanton in the sequel.} and its orientifold image
wrapping the special Lagrangian three-cycle $\Xi$ and $\Xi'$,
respectively. For simplicity we consider situations with an
intersection pattern of the type \bea [\Xi'\cap \Xi]^+ = n^+ =
[\Pi_{{\rm O}6}\cap \Xi]^+, \quad\quad  [\Xi'\cap \Xi]^- = n^- =
[\Pi_{{\rm O}6}\cap \Xi]^- . \eea After identification of the zero
modes from open strings in the $E-E$ and $E'-E'$ sector, the
universal zero modes comprise the four bosonic modes $x^{\mu}$ and
their fermionic partners $\theta^{\alpha}$ and $\ov\tau^{\dot
\alpha}$. To avoid complications due to deformation zero modes, we
assume $\Xi$ and $\Xi'$ to be rigid.

As we can read off from table \ref{antizero}, additional zero modes
arise in the $E-E'$ sector (see \cite{Blumenhagen:2007bn} for a
derivation). Positive intersections give rise to the bosonic modes
$m_{E' E}$, $\ov m_{E E'}$ and the anti-chiral fermion $\ov
\mu^{\dot \alpha}_{E E'}$. The  chiral fermionic modes
$\mu^{\alpha}_{E' E}$ are projected out by the orientifold action. Negative intersections yield the
corresponding modes in the conjugate representation, i.e. bosonic
modes $n_{E E'}$, $\ov n_{E' E}$ and anti-chiral fermion  $\ov
\nu^{\dot \alpha}_{E' E}$. Note that the bosons $m$ and $n$
correspond to the recombination moduli $\phi_i^+$ and  $\phi_j^-$ in
the notation of equ. (\ref{D-term0}).
\begin{table}[h]
\centering
\begin{tabular}{|c|c|c|}
\hline
 zero mode & $(Q_E)_{Q_{ws}}$ &   Multiplicity \\
\hline \hline
 $m, \ov m$ & $(2)_1$ ,$(-2)_{-1}$ & ${1\over 2}[ \Xi'\cap \Xi+\Pi_{{\rm O}6}
\cap \Xi  ]^+$  \\
 $\ov\mu^{\dot \alpha}$ & $(-2)_{1/2}$ & ${1\over 2} [\Xi'\cap \Xi+\Pi_{{\rm O}6}
\cap \Xi]^+$  \\
 $\mu^{\alpha}$ & $(2)_{-1/2}$ & ${1\over 2}[\Xi'\cap \Xi-\Pi_{{\rm O}6}
\cap \Xi]^+$  \\
\hline
$n, \ov n$ & $(-2)_1$ ,$(2)_{-1}$ & ${1\over 2}[\Xi'\cap \Xi+\Pi_{{\rm O}6}
\cap \Xi]^-$  \\
 $\ov\nu^{\dot \alpha}$ & $(2)_{1/2}$ & ${1\over 2}[\Xi'\cap \Xi+\Pi_{{\rm O}6}
\cap \Xi]^-$  \\
 $\nu^{\alpha}$ & $(-2)_{-1/2}$ & ${1\over 2}[\Xi'\cap \Xi-\Pi_{{\rm O}6}
\cap \Xi]^-$  \\
\hline
\end{tabular}
\caption{Charged zero modes at an $E2-E2'$ intersection. 
\label{antizero} }
\end{table}\vspace{5pt}

{\bf{ a) vector-like intersections}} \vspace{5pt}\\

 Vector-like intersections of type $n^+=n^-=1$ were analysed in \cite{Blumenhagen:2007bn} and \cite{GarciaEtxebarria:2007zv}.
As found in \cite{Blumenhagen:2007bn}, the two extra Goldstone modes
$\ov \tau^{\dot \alpha}$ are in fact lifted through couplings in the
instanton effective action of the type $ m \, \ov \mu^{\dot \alpha}
\, \ov \tau_{\dot \alpha} - n \, \ov \nu^{\dot \alpha} \, \ov
\tau_{\dot \alpha}$. Without additional couplings that also lift the
orthogonal combination of fermionic zero modes $\ov \mu^{\dot
\alpha}$ and $\ov \nu^{\dot \alpha}$, the $E-E'$ system contributes
at best to higher fermionic F-terms. This is the situation e.g. for
rigid factorisable three-cycles on toroidal orbifolds, where CFT
computations show that no lifting terms of the required form are
present. More generally, there can exist couplings in the instanton
effective action of the type $(M N)^2$
\cite{GarciaEtxebarria:2007zv}, where $M$ and $N$ formally denote
chiral superfields with the above bosonic and fermionic components.
These can lift the additional fermionic modes and induce
superpotential contributions. Couplings of this type can be viewed
as effective couplings derived from trilinear interactions $M \Phi
N$, where $\Phi$ denotes a massive adjoint superfield corresponding
to a deformation modulus of the wrapped cycle which is obstructed at
second order. For non-zero, but finite mass of the adjoint
integrating $\Phi$ out results in the above quartic coupling which
is suppressed by the inverse $({\rm mass})^2$. In this sense,
absence of the above couplings at the orbifold point reflects the
fact that for rigid cycles, all would-be adjoint scalars are
projected out by the orbifold action so that their mass formally is
$\infty$.

Recall from section \ref{sec_boundstate1} that vector-like intersections are very special in that on both sides of the line of marginal stability BPS objects with charge $[E] + [E']$ exist, at least in a local neighbourhood.
The respective BPS states $E \# E'$ and $E' \# E$ correct the superpotential on either side if and only if the $E-E'$ system does so on top of the line of marginal stability \cite{GarciaEtxebarria:2007zv}. The presence (or absence) of the above quartic couplings is equivalent to rigidity (or not) of the combined objects $E \# E'$ and $E' \# E$ and can thus be verified geometrically. We will exploit this point further in section \ref{sec_instmod}. \vspace{5pt}\\

{\bf{ b) Chiral intersections}} \vspace{5pt}\\
We now proceed to an analysis of chiral intersections with special emphasis on the question how the superpotential behaves upon decay of BPS instanton bound states across the line of marginal stability. For simplicity we stick to the situation $n^+=1, n^-=0$.
This case was considered in \cite{Blumenhagen:2007bn}. As opposed to
the non-chiral intersection,  the following complication arises: In
a globally defined string vacuum, the string theoretic consistency
conditions enforce the presence of extra charged fermionic zero
modes $\lambda^i$. These correspond to open strings between the
instanton and one of the $D6$-branes present in the model
\cite{Ganor:1996pe,Blumenhagen:2006xt,Ibanez:2006da,Florea:2006si}.
These charged zero modes will be called chiral excess modes in the
sequel as they cancel the excess of $U(1)_E$ charge in the instanton
measure arising from the  modes $\ov \mu^{\dot \alpha}$, whose CPT
conjugated counterparts $\mu^{\alpha}$ are projected out. Indeed,
the tadpole cancellation condition can be used to show that the net
total charge of such zero modes adds up to \bea
\sum_i Q_E(\lambda^i) = 
- \sum_a N_a \,  \Xi \circ (\Pi_a + \Pi_{a'}) = - 4 \,\, \Xi
\circ \Pi_{O6} = 4. \eea
More details can be found in \cite{Blumenhagen:2007bn}.

While there are many situations in agreement with this constraint
conceivable, we assume for simplicity there exists a single D6-brane
wrapping some orientifold invariant\footnote{The assumption that
$\Pi_a$ is invariant is not essential, and more general
configurations are equally possible. In case the D-brane $a$ rather
gives rise to a $Sp(2)$ than to an $O(1)$ gauge theory all
intersection numbers including $a$ need to be divided by 2.} sLag
cycle $\Pi_a =\Pi_{a'}$ with $\Pi_a \circ \Xi = 4$ and corresponding
zero modes $\lambda^i_{aE}$, $i = 1, \dots 4$. Note that each of
these modes are identified with one of the modes in the sector
$E'-A$.

Again, the two extra Goldstone modes $\ov \tau^{\dot \alpha}$ are lifted through couplings in the instanton effective action of the type $ m \, \ov \mu^{\dot \alpha} \, \ov \tau_{\dot \alpha}$.
The crucial question is whether or not one can find couplings in the instanton effective action of $E$ and $E'$ which allow us to integrate out also the charged excess modes $\lambda^i$.
One can convince oneself that perturbatively in $g_s$ no such couplings can exist: The only possibility in agreement with charge conservation would be couplings of the type $\ov m \, \lambda^i \, \lambda^j$ or generalisations thereof containing additional products of open string fields. But due to the different worldsheet chirality of the modes $\lambda^i$ and $\ov m$ couplings of this type vanish, following a classic ${\cal N}=2$ worldsheet argument \cite{Distler:1988ms}. By contrast, all purely chiral combinations of the type $m \, \lambda^i \, \lambda^j$ violate instanton $U(1)_E$ charge.
It was concluded in \cite{Blumenhagen:2007bn} that a single $U(1)$ instanton pair of this chiral type cannot contribute to the superpotential.

\subsection{Non-perturbative lifting of charged zero modes}
\label{sec_NPlifting1}

By contrast, it might well happen   that the charged excess modes are lifted through the interaction with other D-brane instantons.
In fact, D-brane instantons can induce superpotential couplings in the worldvolume theory of other D6-branes which are forbidden perturbatively \cite{Blumenhagen:2006xt,Haack:2006cy,Ibanez:2006da,Florea:2006si}. The solution to the above problem would then be to invoke such couplings involving the excess modes $\lambda^i$ in the instanton effective action. The result will be a multi-instanton contribution to the superpotential. A related discussion of multi-instanton effects in non-chiral configurations has been given in \cite{GarciaEtxebarria:2007zv}; for a recent treatment of different aspects of multi-instantons see \cite{Blumenhagen:2008ji} and also \cite{Grimm:2007xm}.

In order to avoid the generation of even more charged excess modes
we consider the possible lifting via extra $O(1)$ as opposed to
$U(1)$ instantons. As will become apparent, the simplest possible
such situation involves \emph{two} more $O(1)$ instantons
$\widetilde E_1$ and $\widetilde E_2$ wrapping the invariant cycles
$\widetilde \Xi_1$ and $\widetilde \Xi_2$, respectively, with
non-vanishing intersections being precisely \bea \label{Int1}
[\widetilde \Xi_1 \cap \Pi_a]^{+} = 2 = [\widetilde \Xi_2 \cap
\Pi_a]^+, \quad\quad\quad [\Xi \cap \widetilde \Xi_1]^+ = 1 = [\Xi
\cap \widetilde \Xi_2]^+. \eea The situation is depicted in figure
\ref{multi_fig}. In Appendix \ref{App_IIA} we construct an explicit
example of such a multi-instanton configuration on the toroidal
orbifold $T^6/{\mathbb Z}_2 \times {\mathbb Z}'_2$. Each of the
$O(1)$ instantons contributes, in the universal sector, the
Goldstone modes $\widetilde x_i^{\mu}$ and $\widetilde
\theta_i^{\alpha}$, and to avoid extra deformation modes we assume
the wrapped cycles are rigid. The $\widetilde E_1-D6_a$ and
$\widetilde E_2-D6_a$ sectors yield two charged fermionic zero modes
each, $\widetilde \lambda^i_1$ and $\widetilde \lambda^i_2$. Given
the nature of the cycles $\widetilde \Xi_1, \widetilde \Xi_2, \Pi_a$
as invariant cycles, the intersection is actually vector-like, but
half the modes are projected out, leaving us again with a chiral
spectrum.

There are also modes between the $U(1)$ instanton and the two $O(1)$
instantons, given by $(k_1, \kappa_1^{\alpha})$ and their charge
conjugate $(\ov k_1, \ov \kappa_1^{\dot \alpha})$, and similarly for
$\widetilde E_2$. Note that, in contrast to the $E-E'$ sector, both
the chiral and anti-chiral bosonic and fermionic fields survive the
orientifold projection here as this sector is not invariant under
$\Omega \ov \sigma$.

\begin{table}[h]
\centering
\begin{tabular}{|c|c|c|c|}
\hline
 zero mode & sector & repr.&   multiplicity \\
\hline \hline
$m $                                & $E-E'$   & $(2_E)$  &       $[\Xi'\cap \Xi]^+ =1$  \\
$\ov m$, $\ov\mu^{\dot \alpha}$     & $E-E'$   &   $(-2_{E})$    &  $[\Xi'\cap \Xi]^+ =1$   \\
\hline
$k_1$, $\kappa^{\alpha}_1$          & $\widetilde E_1 - E $   & $(1_{\widetilde E_1},-1_E)$    & $[\Xi \cap \widetilde \Xi_1]^+ =1 $    \\
$\ov k_1$, $\ov \kappa^{\dot \alpha}_1 $          & $\widetilde E_1 - E $   & $(1_{\widetilde E_1},1_E)$ & $[\Xi \cap \widetilde \Xi_1]^+ =1 $    \\
$k_2$, $\kappa^{\alpha}_2$         & $\widetilde E_2 - E $   & $(1_{\widetilde E_2},-1_E)$    & $[\Xi \cap \widetilde \Xi_2]^+ =1 $    \\
$\ov k_2$, $\ov \kappa^{\dot \alpha}_2 $          & $\widetilde E_2 - E $   & $(1_{\widetilde E_2},1_E)$ & $[\Xi \cap \widetilde \Xi_2]^+ =1 $    \\
                 \hline
$\lambda^i$             & $E-D6_a$ & $(1_E,-1_a)$      & $[\Pi_a \cap \Xi]^+ = 4$ \\
$\widetilde \lambda_1^i$& $ \widetilde E_1 - D6_a $ & $(1_{\widetilde E_1},1_a)$ & $ [\widetilde \Xi_1 \cap \Pi_a]^+ = 2 $ \\
$\widetilde \lambda_2^i$& $ \widetilde E_2 - D6_a $ & $(1_{\widetilde E_2},1_a)$ & $ [\widetilde \Xi_2 \cap \Pi_a]^+ = 2 $ \\

\hline
\end{tabular}
\caption{Summary of boundary changing zero modes. 
\label{table_modes2} }
\end{table}

\begin{figure}[h]
\begin{center}
 \includegraphics[width=0.4\textwidth]{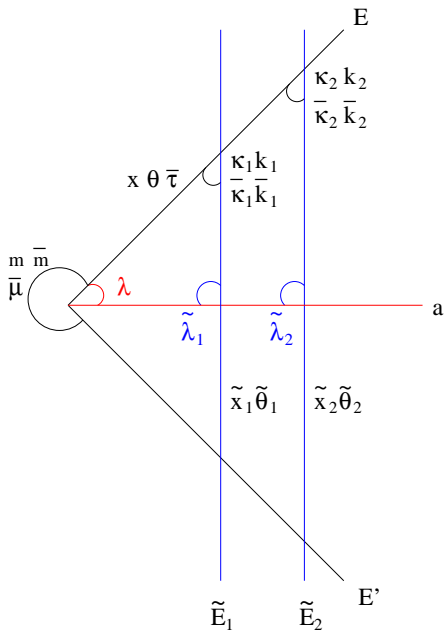}
\end{center}
\caption{\small Multi-instanton configuration involving two O(1)
instantons.}\label{multi_fig}
\end{figure}

We can now analyse the combined instanton effective action involving these fields.
In this section we start on the hypersurface in complex structure moduli space where the $U(1)$ instanton $E$ is supersymmetric with respect to the orientifold plane. On this locus, the bosonic modes are massless. The relevant parts of the effective action of the multi-instanton effective action first include the couplings

\bea \label{S_1} S_1=    Y_{1ij}  \left(   \kappa^{\alpha}_1
\widetilde\theta_{1 \alpha} \, \widetilde\lambda^i_1 \, \lambda^j +
k_1 \widetilde\lambda^i_1 \, \lambda^j \right)   \quad\quad
+\quad\quad (1 \leftrightarrow 2) \eea involving the charged modes
$\lambda^i$ which we are trying to lift. For their computation see \cite{Cvetic:2007ku}.

A second class of couplings can be understood as coming from F-terms
of the type \bea \label{sup1} W \simeq \sum_{i=1}^2   M\, K_i\, K_i,
\eea where $K_i$ formally denotes the superfield associated with the
zero modes $(k_i,\kappa_i^{\alpha}) $ and similarly for $M$
\footnote{Recall, however, that the chiral fermion $\mu^{\alpha}$ is
projected out.}. In components the fermionic terms are \bea S_2=
\label{S_2} {\ov L}_1 \left( \ov\mu^{\dot \alpha} \, \ov
\kappa_{1\dot \alpha} \, \ov k_1  +
  \ov m \, \ov \kappa_{1}^{\dot \alpha} \, \ov \kappa_{1\dot \alpha} \right)+L_1  m \, \kappa_1^{\alpha}\, \kappa_{1 \alpha}
   \quad\quad +\quad\quad (1
 \leftrightarrow
 2),
\eea where we introduced the physical coupling constants $L_1$,
$L_2$. These are related to the holomorphic coupling constants $l_1$
and $l_2$ via \eqref{nh yukawa} described in appendix \ref{App_IIA}.

A third class of interactions consists of the couplings
\cite{Blumenhagen:2007bn} \bea \label{S_3} S_3= C_m\,  (m \, \ov
\mu^{\dot\alpha} \, \ov \tau_{\dot\alpha}) + C_{k_1} \,
(\ov\kappa_1^{\dot \alpha} \, \ov \tau_{\dot \alpha} \,  k_1) +
C_{k_2} \, (\ov\kappa_2^{\dot \alpha} \, \ov \tau_{\dot \alpha} \,
k_2). \eea The bosonic fields furthermore enter the D-term for
$U(1)_E$ in the usual way as \bea \label{S_D} S_D = \frac{1}{2
g^2_E}\, (2 m \ov m - k_1 \ov k_1 -k_2 \ov k_2 - \xi)^2, \eea where
the gauge coupling of the instanton theory $\frac{1}{g^2_E} =
\frac{1}{g_s} \frac{{\rm Vol}_{E2}}{\ell_s^3}$ induces an inverse
scaling with $g_s$, as will become crucial later on\footnote{The
normalisation of the D-term is chosen such that the kinetic terms
for all instanton modes scale as $\frac{1}{2 g^2_E}$. For conventions and their consequences for the vertex
operators see \cite{Cvetic:2007ku}.}.
Besides, the F-term potential associated with the above trilinear
couplings reads\footnote{We thank Ofer Aharony for discussions in the course of which a mistake in an earlier version was noticed.}
\bea
S_F = l_1^2 |m\, k_1| ^2 + l_2^2 |m\, k_2| ^2 + |l_1 k_1^2 + l_2 k_2^2| ^2.
\eea

With the help of the above coupling terms we can indeed saturate all
fermionic zero modes other than the universal $\theta^{\alpha}$
required for superpotential contributions of $E$. Concretely, we
pull down  \bea \label{coup0} Y_{1ij}\,Y_{2kl} \ \times
\,\quad ( \kappa^{\alpha}_1 \widetilde\theta_{1 \alpha} \,
\widetilde\lambda^i_1 \, \lambda^j) \,\,\,   (\kappa^{\alpha}_2
\widetilde\theta_{2 \alpha} \, \widetilde\lambda^k_2 \, \lambda^l)
\eea with $i \neq k$ and $j \neq l$ in the instanton path
integral. The remaining fermionic modes can be absorbed by the
product \bea \label{coup1} C_{k_1} \, C_{k_2} \, \ov L_1 \, \ov L_2
\, \, (k_1 \,\ov \kappa_1 \, \ov \tau)  \,\, (\ov \mu \,\ov \kappa_1
\, \ov k_1) \,\,   (k_2 \,\ov \kappa_2 \, \ov \tau) \, \, (\ov \mu
\, \ov \kappa_2 \, \ov k_2). \eea

Schematically, we are left with the nonvanishing, finite bosonic integral
\bea
\label{bos1}
 Y^2_{1ij}\,Y^2_{2kl} \,\ov L_1\, \ov L_2
\,\int dk_1 \, d\ov k_1 \, \, dk_2 \, d\ov k_2 \, \, dm \, d\ov m \,
\,\,\,   |k_1|^2 \, |k_2|^2 \, \, \,\,  exp(-S_D - S_F). \eea
Instead of (\ref{coup1}) we can also saturate the remaining
fermionic modes by \bea  (C_m)^2 \, \ov L_1 \, \ov L_2 \,\, (m \,
\ov \mu \, \ov \tau)^2 \, \, \ov m \, \ov \kappa_{1} \, \ov \kappa_1
\, \, \ov m \, \ov \kappa_{2} \, \ov \kappa_2, \label{coup2} \eea
which leads to the non-vanishing bosonic integral \bea \label{bos2}
Y^2_{1ij}\,Y^2_{2kl}\, \ov L_1 \,\ov L_2\,\int dk_1 \, d\ov k_1 \,
\, dk_2 \, d\ov k_2 \, \, dm \, d\ov m \, \,\,\, |m|^4 \, \, \,\,
exp(-S_D - S_F). \eea

As a result of summing up all different channels, the
multi-instanton BPS configuration produces a non-vanishing
contribution to the superpotential. The scale of this contribution
is set by the exponentiated classical instanton action, \bea
\label{Omegadep1} W \simeq exp \left(- \frac{2 \pi}{\ell_s^3}
(\int_{\Xi} \frac{1}{g_s}  \Omega + i C_3 + \int_{\widetilde \Xi_1}
\frac{1}{ g_s}  \Omega + i C_3 + \int_{\widetilde \Xi_2} \frac{1}{
g_s}  \Omega + i C_3 ) \right). \eea

As in single instanton computations, this classical suppression
factor is multiplied by the exponentiated sum over all one-loop
annulus diagrams with one end on the instantons and one end on the
D6-branes of the model, $\sum_b Z'_A(E2, D6_b)$, together with the
M\"obius amplitudes $M'(E2, O6)$ \cite{Blumenhagen:2006xt}. Here
$E2= E, \widetilde E_1, \widetilde E_2$ and the  massless modes are
excluded. As an important consistency check, holomorphicity of the
generated superpotential is ensured by the cancellation of the
non-holomorphicities in the physical couplings $\ov L_i$, $Y_{1
jk}$, $Y_{2 jk}$ appearing in (\ref{coup0}), (\ref{coup1}),
(\ref{coup2}), partially among one another and partially with the
non-holomorphic part of these one-loop amplitudes. More details are
given in the context of our concrete example at the end of appendix
\ref{App_IIA}.

Before proceeding we would like to notice that the simpler
configuration consisting of the $U(1)$ instanton pair and only one
$O(1)$ instanton does not induce a superpotential. While for
suitable intersection numbers the resulting effective action may
contain the couplings required to saturate all extra fermionic zero
modes, the complex integral over the bosonic modes contains now a
monomial in $m^a k^b$ and not in $|m|^a |k|^b$. It vanishes as a result of the
uncancelled relative phase. We will come back to this point at the end of the next section.

\subsection{(Non-)BPS bound states and contributions to the superpotential}
\label{fayet}
\noindent I.) $\xi > 0$ \\

Now we deform the complex structure of the Calabi-Yau manifold
away from the line of marginal stability ${\cal M}_0$ determined by $\xi =0$ for the cycles $\Xi$ and $\Xi'$.
For simplicity we assume we can take a path in complex moduli space along which the calibration of the other D-branes remains unchanged \footnote{This is not implying a continuous change of moduli, but is rather meant as a gedanken experiment to analyse the system for different values of the complex structure moduli. 
In general the closed string  moduli may possess a non-trivial potential instead of being free parameters. In particular, the instanton under consideration induces a complex structure moduli dependence of the potential via its exponential. The microscopic lifting of zero modes does not depend, however, on this backreaction of the instanton on the geometry. For example if the instanton induced coupling involves products of open string fields, the complex structure moduli will in general not be fixed by the instanton sector.}.

Due to the strictly chiral nature of the intersection of the cycle $\Xi$ with its image $\Xi'$, (\ref{Int1}), it is possible only for deformations into $\cal M_+$ where $\xi > 0$  that $\Xi$ and $\Xi'$ combine into a new special Lagrangian cycle
\bea
Y = E' \# E
\eea
with homological charge $[E]+[E']$ and which preserves the same ${\cal N}=1$ supersymmetry as the orientifold. The bound state $Y$ disappears from the spectrum of BPS branes on the other side in complex moduli space, i.e in ${\cal M}_-$. It is therefore an interesting question how the instanton-induced superpotential behaves as the line of marginal stability is crossed.

Let us begin with small deformations leading to formation of the BPS
bound state $Y$. From the effective field theory point of view, the
Fayet-Iliopoulos parameter $\xi$ for $U(1)_E$ becomes positive and
renders the bosons $m, \ov m$ tachyonic. At the end of the
recombination process $m, \ov m$ have acquired a VEV such that
D-flatness is preserved. The fluctuation modes $\delta m$ and $
\delta \ov m$ become massive via the D-term, and so do the fermions
$\ov \mu^{\dot\alpha}$ and $\ov \tau^{\dot \alpha}$ through the
coupling $\langle m\rangle \, \ov \mu \, \ov \tau$. The VEV for $m$
and  $\ov m$ likewise induces a mass term for the bosonic modes
$k_i, \ov k_i$ and the fermions $\kappa_i^{\alpha}$ and $\ov
\kappa_i^{\dot\alpha}$.

The only massless modes of the multi-instanton system  (besides $x^{\mu}, \theta^{\alpha}$ and $\widetilde x_i^{\mu}$) are the charged  modes $\lambda^i$, $\widetilde\lambda^j$ together with $\widetilde \theta_1^{\alpha}, \widetilde \theta_2^{\alpha}$.
From general ${\cal N}=2$ worldsheet arguments there should exist the six-point couplings
\bea
\label{6point}
\langle \widetilde \theta_1^{\alpha} \, \widetilde \theta_1^{\beta}  \, \, \widetilde\lambda^i  \lambda^j \widetilde\lambda^k \lambda^l \rangle + 1 \leftrightarrow 2.
\eea
The easiest way to see this is to put the vertex operators for the respective zero modes in the following pictures,
\bea
\langle V^{1/2}_{1/2}(\widetilde \theta_1^{\alpha})\,\,  V^{-1/2}_{3/2}(\widetilde \theta_1^{\beta}) \,\,\,
V^{-1/2}_{-1/2}(\widetilde \lambda^i)  \,\,\,
V^{-1/2}_{-1/2}(\lambda^j) \,\,\,
V^{-1/2}_{-1/2}(\widetilde \lambda^k)   \,\,\,
V^{-1/2}_{-1/2}( \lambda^l)    \rangle,
\eea
where the superscript denotes the ghost picture and the subscript the worldsheet $U(1)$ charge.
Pulling down these two couplings therefore saturates all extra
fermionic modes, and the instanton bound state $(E' \#E) \cup \,
\widetilde E_1 \cup \, \widetilde E_2$ contributes to the
superpotential.

There is an alternative way to describe the system by thinking of the instantons wrapping the individual cycles $\Xi, \Xi', \widetilde \Xi_1$ and $\widetilde \Xi_2$ before formation of the bound state $Y$ in the following way: As $\Xi$ and $\Xi'$ are at non-supersymmetric angles, the open string excitations describing the bosons $m, \ov m$ are tachyonic, while the ones corresponding to  $k_i, \ov k_i$ acquire positive ${\rm (mass)}^2$. From the quantisation of the open string modes it is furthermore clear that in this picture, i.e. prior to condensation of $m, \ov m$, all fermionic modes remain massless.
The instanton effective action for this system is obtained by integrating out the bosonic non-zero modes and keeping only the couplings involving the fermionic zero modes. In fact, non-zero instanton modes are strictly off-shell as it is not possible, in absence of four-dimensional momentum, to write down a consistent vertex operator for massive excitations. This is reflected in the usual procedure to allow for the non-zero modes to appear only in the one-loop amplitudes.

The effective coupling replacing the interactions (\ref{S_2}) and (\ref{S_3})  upon integrating out $k_i$ and $\ov k_i$ become
\bea
S' = \ov \mu \, \ov \kappa_1 \, \ov \kappa_1 \, \ov \tau + \ov \mu \, \ov \kappa_2 \, \ov \kappa_2 \, \ov \tau.
\eea

These terms allow us to saturate all extra fermionic zero modes, reproducing the conclusion that the instanton system contributes to the superpotential.
\\

\noindent II.) $\xi < 0$ \\

Now we deform the complex structure such as to enter the region
${\cal M}_-$ of moduli space where the special Lagrangian $Y$ ceases
to exist. However, as encoded already in the D-term potential
(\ref{S_D}), $E$ can recombine instead with the $O(1)$ instantons on
the cycles $\widetilde \Xi_1$ or  $\widetilde \Xi_2$. The D-term
only fixes the combination $|k_1|^2 +
 |k_2|^2$ and leaves us with
one complex bosonic modulus consisting of the orthogonal combination
as well as the relative phase between the complex fields $k_1$ and
$k_2$. Both are fixed by the F-term in a D- and F-flat manner. The
non-zero VEV for $k_1$ and $k_2$ also renders the boson $m$ massive.
All extra fermionic modes $\widetilde\lambda_i$, $\lambda_j$,
$\kappa_k$ and $\ov\kappa_l$ acquire a mass via their couplings to
$k_i$.


This shows how holomorphicity of the D-brane instanton induced superpotential is maintained even in situations where specific BPS instantons disappear across lines of marginal stability. In ${\cal M}_+$ the superpotential is corrected by instantons wrapping the BPS configuration $(E' \#E) \cup \, \widetilde E_1 \cup \, \widetilde E_2$. It is a multi-instanton configuration with constituents $\widetilde E_1$, $\widetilde E_2$ and the BPS bound state $Y=E' \#E$. Along ${\cal M}_0$ this bound state $Y$ meets a line of marginal stability, but the multi-instanton
$E  \cup E'  \cup \, \widetilde E_1 \cup \, \widetilde E_2$
is still BPS and contributes to the superpotential. In ${\cal M}_-$ the former BPS state $Y=E' \#E$ has disappeared, but ther exists a new
BPS state $\Psi = \widetilde E_1 \# (E \cup E') \# \widetilde E_2$ with charge $[E] + [E'] + [\widetilde E_1] + [\widetilde E_1]$.
The two additional instantons $\widetilde E_1$ and $\widetilde E_2$ required to lift the fermionic zero modes for $\xi = 0$ conspire such that the number of BPS states of total charge $[E] + [E'] + [\widetilde E_1] + [\widetilde E_1]$ does not jump across the line of marginal stability. \vspace{10pt}\\

To illustrate this connection further, it is instructive to analyse how a jump in the BPS spectrum is correlated with a microscopic obstruction to a superpotential contribution already at threshold. The simplest example would of course be just the $U(1)$ instanton and its image which cannot contribute due to extra charged modes. But there are even more subtle obstructions to superpotential contributions in agreement with a discontinuous BPS spectrum.

\begin{figure}[h]
\begin{center}
\includegraphics[width=0.35\textwidth]{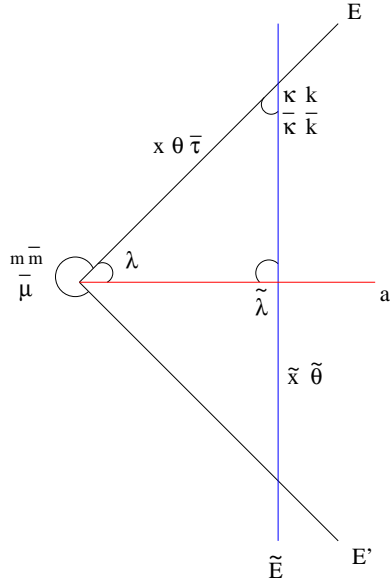}
\end{center}
\caption{\small Multi-instanton configuration involving a single
O(1) instanton.}\label{multi_fig2}
\end{figure}

Consider the lifting of the charged zero modes by a single
$O(1)$ instanton wrapping the cycle $\widetilde{\Xi}$. In order to
lift the additional 4 charged zero modes $\lambda$ we
require e.g. \footnote{Our results hold also true for different
intersections in the $E-\widetilde{E}$ sector.} \bea \label{Int1}
[\widetilde \Xi_1 \cap \Pi_a]^{+} = 4, \quad\quad\quad [\Xi \cap
\widetilde \Xi_1]^+ = 1. \eea Such a setup is depicted in figure \ref{multi_fig2}. The massless spectrum comprises $4$ additional charged zero modes $ \tilde{\lambda} $
the bosonic and fermionic zero modes $k$ and $\kappa^{\alpha}$
and their conjugates
and finally the universal
zero modes $\tilde{x}$ and $\tilde{\theta}^{\alpha}$ of the $O(1)$
instanton.

One observes the same couplings as in (\ref{S_1}), (\ref{S_2}) and
(\ref{S_3}), but the D-term and the F-term  now take the form \bea
\label{S_D} S_D = \frac{1}{2 g^2_E}\, (2 m \ov m - k \ov k-
\xi)^2,\qquad S_F = l_1^2 \left( (k \,  \ov k) ^2 +  |m \, k| ^2
\right). \eea As before we can saturate all
charged zero modes $\lambda$ and $\tilde{\lambda}$ by pulling down
 \bea \label{1coup0} Y_{ij}\,Y_{kl} \ \times
\,\quad ( \kappa^{\alpha} \widetilde\theta_{\alpha} \,
\widetilde\lambda^i \, \lambda^j) \,\,\,   (k \,\widetilde\lambda^k
\, \lambda^l), \eea while the remaining fermionic zero modes can be
absorbed by \bea C_{k} \,C_m \, \ov L \, \, (k \,\ov \kappa \, \ov
\tau) \,\, (\ov \mu \,\ov \kappa \, \ov k) \,\,(\ov \mu \, m \, \ov
\tau). \eea This leaves us with the bosonic integral \bea \label{1bos1}
 Y^4_{ij} \, C_k \,\ov L
\,\int dk \, d\ov k \, \, dm \, d\ov m \, \,\,\,   |k|^2 \, k^2
\, m\, \, \,\,  exp(-S_D - S_F). \eea Unlike in the previous case this vanishes after integrating over the
relative phase between $m$ and $k$. Alternatively one can saturate
all fermionic zero modes via the couplings \bea C^2_{m} \,  \ov L \,
\, (\ov m \,\ov \kappa \, \ov \kappa) \,\,\, (\ov \mu \, m \, \ov
\tau)^2 \eea leading to the bosonic integral \bea \label{2bos1}
 Y^4_{ij} \, C^2_m \,\ov L
\,\int dk \, d\ov k \, \, dm \, d\ov m \, \,\,\,   |m|^2 \, k^2
\, m\, \, \,\,  exp(-S_D - S_F). \eea Again this integral  vanishes and there are no superpotential
contributions.



This is consistent with the behaviour of such an instanton configuration for a
small deformation of the complex structure.
The configuration is very similar to the D-brane setup discussed
in section \ref{sec_F-term_obst}.
For sufficiently small deformations of the complex structure
the new stable
geometric object is described by condensation of the bosonic modes
such that the potential $V = S_D + \, S_F$  is minimized. For $\xi < 0$, this happens at \bea |k| =
\sqrt{-\frac{\xi}{1+a}}, \quad \quad m=0, \eea where $a = 2 g^2_E \,
l <<1$.  Note that this minmum breaks both D-flatness
and F-flatness. It corresponds to an instanton wrapping the bound
state $\widetilde \Psi$ of the cycle $\Xi \cup \Xi'$ with
$\widetilde \Xi$.

This new multi-bound state $\widetilde \Psi$ is truly non-BPS. As in
section \ref{sec_F-term_obst} a possible way to think about
$\widetilde \Psi$ is as a deformation of the sLag $\Psi$ defined as
the would-be BPS bound state formed by $\widetilde \Xi$, $\Xi$ and
$\Xi'$ if the superpotential (\ref{sup1}) were absent, i.e. $l=0$.
From the
field theory point of view, the tachyon $k$  would
condense as $|k| = \sqrt{-{\xi}}$ and the excitation mode
$\delta k$ around this vacuum expectation value would be massive.
Now by switching on $l \neq 0$, $m$ acquires
a mass.  The F-terms also induce a term  linear in the massive fields
$\delta k$. This indicates that the system is unstable towards
formation of the metastable non-BPS state $\widetilde \Psi$.
In the spirit of the discussion at the end of section
\ref{sec_F-term_obst}, $\widetilde \Psi$ is a \emph{non-calibrated Lagrangian
three-cycle}.

Such a
non-supersymmetric state is not expected to contribute to the
superpotential, thus on the line of margin stability $\xi=0$ there
should not be any contributions either. This is in complete
agreement with our previous analysis.

\section{Superpotential contributions in Type I theory}
\label{sec_TypeI}

In this section we give the mirror dual description of instanton bound states for Type I compactifications  on a Calabi-Yau manifold $X$.
This language is particularly useful to illustrate the general ideas of section \ref{MultiIIA} in terms of very concrete and computable algebraic objects. By translating the results of the previous section into Type I we will identify a new class of instantons correcting the superpotential. By S-duality to heterotic compactifications they map to bound states of worldsheet and NS5-brane instantons carrying vector-bundles.

The building blocks of the Type I gauge sector are formed by stacks
of $M_a = N_a \times n_a$ $D9$-branes wrapping $X$ and carrying
stable holomorphic vector bundles $V_a$ of rank $n_a$. In addition
we have to add their orientifold images, given by a $D9$-brane with
the dual bundle $V_a^{\vee}$. The associated gauge group of each
stack of $D9$-branes is $U(N_a)$. For pairs of such magnetised
$D9$-branes, bifundamental  matter is counted by the cohomology
groups \bea H^i(X,V_a \otimes V_b), \quad\quad i =1,2, \eea where
$i=1$ and $i=2$ respectively refer to chiral and anti-chiral
superfields transforming as $(N_a,N_b)$. Replacing $V_a$ by its dual
$V^{\vee}_a$ interchanges $N_a$ with the conjugate representation
$\ov N_a$. More background can be found in
\cite{Blumenhagen:2005pm,Blumenhagen:2005zh}. We will be working in
the large volume limit where the BPS condition for the holomorphic
bundles is given, in slight oversimplification, by $\mu$-stability
together with \bea \label{muslope} \mu_J(V) =0, \quad\quad\quad
\mu_J(V) = \frac{1}{{\rm rk}(V)} \int_X J \wedge J \wedge c_1(V).
\eea
Strictly speaking, BPS bundles are not described by the category of coherent sheaves, but rather by its derived category \cite{Douglas:2000gi}. The correct stability criterion differs from the above even in the limit $\alpha' \rightarrow 0$, where perturbative and worldsheet instanton corrections to the definition of the $\mu$-slope can be neglected \cite{Aspinwall:2004jr,Diaconescu:2007bf}.

The superpotential receives corrections from Euclidean $D1$-branes wrapping holomorphic curves \cite{Witten:1999eg}. These are dual to the described $O(1)$ instantons in Type IIA. In the sequel it will be useful to model an $E 1$ instanton wrapping the curve $C$ as the sheaf ${\cal O}|_C$.
For a detailed description of $E 1$-instantons in this language we refer the reader to \cite{Cvetic:2007qj}.

One of the motivations for this work was to investigate whether the
superpotential  also receives contributions from  Euclidean
$D5$-branes on $X$. Such $E5$-instantons without additional gauge
flux carry gauge group $Sp(2)$ and therefore exhibit too many
Goldstone modes to contribute to the superpotential at least in a
straightforward manner. Instead we consider $E5$-instantons with
non-vanishing worldvolume flux. Note that these are dual in Type IIA theory to $E2$-instantons of $U(1)$ type, which can meet lines of marginal or threshold stability and were considered in the previous sections.

Similarly to the IIA context, we begin with a configuration of BPS
$E5$-instantons carrying stable holomorphic bundles with zero slope,
together with their orientifold image. For practical reasons we will
mostly focus on $E5$-branes endowed with complex line bundles $L$
(together with their image $E5$-branes with $L^{\vee}$). This can be
generalised to  bundles of higher rank.

We are interested in studying the transition of this system, i.e. of the direct sum of instanton bundles
\bea
\label{Vtilde}
\widetilde V= L \oplus L^{\vee},
\eea
to non-split extension bundles upon crossing a line of marginal stability in K\"ahler moduli space.
The two possible bound states  $V=L^{\vee} \#L$  and  $U=L \# L^{\vee}$ can be thought of as the extensions given by
\bea
\label{ExtV}
0 \rightarrow L \rightarrow V \rightarrow L^{\vee} \rightarrow 0
\eea
and
\bea
\label{ExtV2}
0 \rightarrow L^{\vee} \rightarrow U   \rightarrow L \rightarrow 0,
\eea
respectively. Non-splitness and thus existence of the extensions $V$ or $U$ requires that the groups ${\rm Ext}^1_X(L^{\vee},L)=H^1(X,L^2)$ or ${\rm Ext}^1_X(L,L^{\vee})=H^1(X,(L^{\vee})^2)$ are non-zero, respectively.
A necessary condition for stability of a non-split extension is that the slope of the bundle to the left be smaller than that of the extension bundle. In general this is not sufficient yet, but for small enough deformations away from ${\cal M}_0$, i.e. for sufficiently small slope, stability is expected on physical grounds (see also \cite{Thomas:2001ve}).
The recombination (or extension) modes $m, \ov m$ or $n, \ov n$ in the $L-L^{\vee}$ sector triggering formation of $V$ or $U$, respectively, are summarized in table \ref{extmodes}.

\begin{table}[h]
\centering
\begin{tabular}{|c|c|c|}
\hline
 zero mode & $Q_E$ &  Cohomology \\
\hline \hline
 $m$                            & $2$  & $H^1(X, L^2) $  \\
\hline
$\ov m$, $\ov\mu^{\dot \alpha}$ & $-2$ & $ H^2(X, (L^{\vee})^2) $  \\
\hline
 $n$                            & $-2$  & $H^2(X, L^2) $  \\
\hline
$\ov n$, $\ov\nu^{\dot \alpha}$ & $2$ & $ H^1(X, (L^{\vee})^2) $  \\
\hline
\end{tabular}
\caption{Extension modes in $L-L^{\vee}$ sector.} 
\label{extmodes}
\end{table}\vspace{5pt}

The direct sum $\widetilde V$ represents a BPS instanton along the
real codimension 1 hypersurface ${\cal M}_0$ in K\"ahler moduli
space defined by $\mu_{J}(L) = 0 = \mu_{J}(\widetilde V)$. Upon
deforming $J$ such that $\mu_{J}(L) < 0$ we enter into ${\cal M}_+$
where the bound state $V$ forms, while $U$ can exists for  $J \in
{\cal M}_-$ defined by $\mu_{J}(L) > 0$.

In this language it is particularly obvious that objects of a given charge vector can meet several lines of marginal/threshold stability in moduli space and that the type of these hypersurfaces can  vary.
In our case this corresponds to the existence of two different line bundles $L_1$, $L_2$ with
\bea
{\rm ch}(L_1) + {\rm ch}(L_1^{\vee}) = {\rm ch}(V) ={\rm ch}(L_2) + {\rm ch}(L_2^{\vee}), \quad H^1(X,L_1^2) \neq 0 \neq H^1(X,L_2^2).
\eea
Since ${\rm ch}_i(L^{\vee}) = (-1)^i \, {\rm ch}_i(L)$ this only constrains ${\rm ch}_2(L_1) ={\rm ch}_2(L_2) = \frac{1}{2} {\rm ch}_2(V)$. By contrast, both the slope of $L_i$ and the index $\chi(L_i^2)$ depend on the odd Chern classes. It can therefore happen that $V$ meets a line of threshold stability with respect to $L_1, L_1^{\vee}$ for K\"ahler class $J_1$ and a line of marginal stability with respect to $L_2, L_2^{\vee}$ for a different K\"ahler class $J_2$.

\subsection{Lifting of chiral excess modes in a 2-instanton process}
\label{sec_2-inst_TypeI}

Consider now a tadpole free supersymmetric Type I compactification
of the above type on the locus $J \in {\cal M}_0$. As in Type IIA
one can show that the charged zero modes of the BPS $E5$-instanton
carrying the direct sum bundle $\widetilde V$ with all $D9$-branes
have total $U(1)_E$ charge \bea \label{sumE} \sum_a Q_a = - 4 \,
\chi(X,L^2). \eea This enforces the existence of chiral excess modes
for non vector-like situations, i.e. whenever $\chi(X,L^2)\neq 0$
\footnote{This is not in conflict with the previous statement about
the change of the intersection type of several lines of
marginal/threshold stability in moduli space. The definition of a
quantum number $U(1)_E$ only makes sense on top of a line of
marginal stability for a chiral intersection as otherwise the BPS
object is actually of $O(1)$ type and the charge of the zero modes
trivially adds up to zero.}. As before we turn to the simplest
chiral case of
 $h^1(X,L^2) = n^+ = 1, h^2(X,L^2) =n^- =0$.  Equ. (\ref{sumE}) can be satisfied
by numerous possible configurations all of which lead to similar conclusions.

For concreteness consider the case that there exists a single $D9$-brane with line bundle $W$ (together with its orientifold image with $W^{\vee}$) such that
\bea
\label{H(WxL)}
&&\lambda^i \in  H^1(X, W \otimes L) \simeq {\mathbb C}^{n+2},      \,\,          H^2(X, W \otimes L) \simeq {\mathbb C}^n, \\
&&\lambda^j  \in H^1(X, W^{\vee} \otimes L) \simeq {\mathbb C}^{n+2},   \,\,  H^2(X, W^{\vee} \otimes L) \simeq {\mathbb C}^n,  \nonumber
\eea
in agreement with (\ref{sumE}).
Let us focus on the minimal case with $n=0$ where have precisely two modes $\lambda^i$ with charges $(1_W,1_E)$ and two $\lambda^j$ in $(-1_W,1_E)$.
In the Type IIA dual we saw from general worldsheet arguments that no perturbative couplings in the instanton effective action can lift these chiral excess modes.
In the present context such couplings for the system $\widetilde V$  on top of the line of marginal stability are forbidden by general properties of the chiral ring structure.
The present formalism allows us to follow the fate of these modes upon formation of the bound state $V$. As detailed in appendix \ref{app_Lifting},
they necessarily survive as vector-like modes of the new bound state.
Clearly the two statements are equivalent as the excess modes are expected to be lifted perturbatively in the recombined system precisely if there exist couplings to the recombination moduli who acquire a VEV upon recombination.

On the other hand, the lifting of the chiral excess modes via two more $O(1)$ instantons is possible in a manner totally analogous to the IIA picture, so that we can be brief.
We need two more such $E1$-instantons wrapping the rigid holomorphic curves $C_1$ and $C_2$ with a charged zero mode
 spectrum as given in table \ref{chargedmodes2}. The cohomology groups follow from the general discussion in \cite{Cvetic:2007qj}.

\begin{table}[h]
\centering
\begin{tabular}{|c|c|c|c|}
\hline
 zero mode & charge & number &  Cohomology \\
\hline \hline
$\lambda^i$                &  $(1_W,1_E) $    & $2$ & $H^1(X, W \otimes L)$   \\
$\lambda^j$                &  $(-1_W,1_E) $   & $2$ & $H^1(X, W^{\vee} \otimes L)$  \\
\hline
$\widetilde \lambda_1^i$    &  $-1_W$ & $2$  & $H^0(C_1, W^{\vee}|_{C_1}(-1)) $  \\
$\widetilde \lambda_2^j$    &  $1_W$ & $2$  & $H^0(C_2, W|_{C_2}(-1)) $  \\
\hline
$k_i, \kappa_i^{\alpha}$            &   $-1_E$  & $1$  & $ H^0(C_i, L^{\vee}|_{C_i}(-1))$ \\
$\ov k_i, \ov \kappa_i^{\dot \alpha}$ & $1_E$   & $1$  & $ H^1(C_i, L|_{C_i}(-1))$ \\
\hline
\end{tabular}
\caption{Charged zero modes.} 
\label{chargedmodes2}
\end{table}\vspace{5pt}
From Bott's theorem applied to the $\IP^{1}$s $C_1, C_2$ this spectrum requires that
\bea
&&W|_{C_1} = {\cal O}_{C_1}(-2), \quad\quad   W|_{C_2} = {\cal O}_{C_2}(2), \nonumber\\
&&L|_{C_i} = {\cal O}_{C_i} (-1), \,\, i=1,2.
\eea

The couplings we invoke to lift all extra fermionic zero modes are as in the IIA system.
E.g. it is possible to lift the excess modes $\lambda \in H^1(X, W \otimes L)$ through couplings of the form
\bea
\label{l_wlift}
\langle \kappa^{\alpha} \, (\theta_{E1})_{\alpha} \, \lambda \, \widetilde \lambda \rangle.
\eea
This Yukawa coupling corresponds to the map
\bea
\label{YukIa}
H^0(C_1, L^{\vee}|_{C_1}(-1)) \otimes   H^0(C_1, W^{\vee}|_{C_1}(-1)) \otimes    H^1(C_1, W \otimes L|_{C_1}) \rightarrow {\mathbb C},
\eea
which is just the pairing
\bea
H^0(C_1, {\cal O} )   \otimes  H^0(C_1, {\cal O}(1) )  \longrightarrow   H^0(C_1, {\cal O}(1) ).
\eea
Note that in (\ref{YukIa}) only those modes $\lambda \in H^1(X, W \otimes L)$   contained in the group  $ H^1(C_1, W \otimes L|_{C_1})$  can couple to $\kappa^{\alpha} \, (\theta_{E1})_{\alpha}$ and $\widetilde \lambda^i$, which are localised at $C_1$. It is therefore to be checked in concrete examples that all $\lambda^i$ are indeed lifted.

Similarly, the analogue of the coupling (\ref{S_2}),
$\langle \ov k_i \, \ov\mu_{\dot \alpha} \, \ov \kappa^{\dot \alpha}_i  \rangle
$,
which is the CPT conjugate version of the Yukawa coupling $M K_i K_i$, see equ. (\ref{sup1}),
corresponds to the map
\bea
 H^0(C_i, L^{\vee}|_{C_i}(-1)) \otimes H^1(C_i,L^2|_{C_i}) \otimes  H^0(C_i, L^{\vee}|_{C_i}(-1)) \rightarrow {\mathbb C}.
\eea
Due to the localisation of the field $K_i$ on $C_i$, $K_i \in H^0(C_i, L^{\vee}|_{C_i}(-1))$,
only the restriction
 $H^1(C_i,L^2|_{C_i})$ can participate in Yukawa couplings.

In situations where all required couplings are non-zero
the multi-instanton configuration on top of the line of marginal stability, $J \in {\cal M}_0$, yields
 a non-vanishing superpotential contribution.
The same conclusion holds for deformations away from ${\cal M}_0$ into ${\cal M}_+$ or ${\cal M}_-$ .
E.g. we propose that for $J \in {\cal M}_-$ the BPS instanton formed by
$\widetilde V$, $C_1$ and $C_2$
contributes to the superpotential.
The relevant BPS state $\Psi$ is
the bound state formed by the skyscraper sheaves ${\cal O}|_{C_i}$ and the vector bundle $L \oplus L'$
\bea
&&0 \rightarrow {\cal O}|_{C_1} \rightarrow V_1 \rightarrow L \oplus L' \rightarrow 0, \nonumber \\
&&0 \rightarrow {\cal O}|_{C_2} \rightarrow  \Psi \rightarrow V_1 \rightarrow 0.
\eea

\subsection{Instanton moduli}
\label{sec_instmod}

In this section we take a closer look at vector-like recombination processes associated with lines of threshold stability \cite{Blumenhagen:2007bn, GarciaEtxebarria:2007zv}.
As summarized in section \ref{sec_def}, superpotential contributions of the system $L \oplus L^{\vee}$ at threshold require the presence of quartic superpotential
couplings between the vector-like extension modes $m$ and $n$. After recombination  these couplings lift otherwise massless moduli of the bound state which are inherited from the recombination moduli of the wrong charge that acquire no VEV.
A convenient way to determine whether or not these couplings are present is therefore to compute the moduli space of deformations of an instanton bound state described by the extension of two rigid vector bundles.

For simplicity we consider the special case that $L$ is a line bundle. The bundle moduli of the self-dual vector bundle $V=V^{\vee}$ are counted by $H^1(X, V \otimes V^*)$. For our extension
\bea
0 \rightarrow L \rightarrow V \rightarrow L^{\vee} \rightarrow 0
\eea
$H^i(X, V \otimes V^*)$ is computed from the long exact sequence induced by
\bea
\label{seq_adj}
0 \rightarrow L \otimes V^{\vee} \rightarrow V \otimes V^{\vee} \rightarrow L^{\vee} \otimes V^{\vee} \rightarrow 0.
\eea
In turn, $H^i(X,L \otimes V^{\vee})$ is determined by the long exact sequence induced by
\bea
0 \rightarrow L \otimes L \rightarrow L \otimes V^{\vee} \rightarrow L \otimes L^{\vee} \rightarrow 0.
\eea
This sequence is given by \vspace{5pt}\\
\begin{tabular}{ccccc}
0 & $\rightarrow H^0(X,L \otimes L)$ & $\rightarrow H^0(X, L \otimes V^{\vee})$ &  $\rightarrow H^0(X,L \otimes L^{\vee}) \stackrel{f}{\rightarrow} $ &\\
  & $\rightarrow H^1(X,L \otimes L)$ & $\rightarrow H^1(X, L \otimes V^{\vee})$ &  $\rightarrow H^1(X,L \otimes L^{\vee}) \rightarrow $             &   \\
  & $\rightarrow H^2(X,L \otimes L)$ & $\rightarrow H^2(X, L \otimes V^{\vee})$ &  $\rightarrow H^2(X,L \otimes L^{\vee}) \rightarrow $             &   \\
  & $\rightarrow H^3(X,L \otimes L)$ & $\rightarrow H^3(X, L \otimes V^{\vee})$ &  $\rightarrow H^3(X,L \otimes L^{\vee}) \rightarrow $ &0.
\end{tabular}
\vspace{5pt}\\

Recall that we assume that there exists a hypersurface ${\cal M}_0$ in K\"ahler moduli space where $\mu(L) =0$ and that there exist
small deformations of the K\"ahler form $J$ into the region ${\cal M}_+ $ where  $\mu(L)  < 0 < \mu(L^{\vee})$. This means that $L$ is neither ample nor anti-ample. In addition we assume that the extension $V$ is non-split
and $V$ is stable for $J \in {\cal M}_+$ at least for sufficiently small deformations of $J$ away from ${\cal M}_0$.

As an immediate consequence of these assumptions $H^0(X,L^2)= 0 = H^3(X,L^2)$ and likewise for $(L^{\vee})^2$.\footnote{Recall that $H^0(X,L^2) \neq 0$ would imply
 the existence of a map ${\cal O} \rightarrow L^2$, but since $\mu(L^2) =0$ for $J \in {\cal M}_0$ this would mean ${\cal O} = L^2$. The statement about $H^3(X,L^2)=0$ follows by Serre duality from $H^0(X,(L^{\vee})^2)= 0$.}
Furthermore $V^{\vee} \otimes L$ is
stable and of negative slope (since we are working in the regime $J \in {\cal M}_+$) so  that $H^0(X, L \otimes V^{\vee})=0$.
Finally, the third column just contains $H^*(X,L \otimes L^{\vee}) = H^*(X, {\cal O}) \simeq (\mathbb C,0,0,\mathbb C)$.
The first line therefore implies that the coboundary map $f$ is an injection and thus of maximal rank 1.
It follows that
\bea
&& h^1(X, L \otimes V^{\vee}) = h^1(X,L^2) -1, \quad\quad  h^2(X, L \otimes V^{\vee}) = h^2(X,L^2), \\
&& h^3(X,L \otimes V^{\vee}) = 1 \label{h3_1}.
\eea

The long exact sequence induced by (\ref{seq_adj}) reads \vspace{5pt}\\
\begin{tabular}{ccccc}
0 &    $\rightarrow H^0(X,V^{\vee} \otimes L) $   &  $\rightarrow H^0(X, V \otimes V^{\vee})$ &   $\rightarrow H^0(X,V^{\vee} \otimes L^{\vee})$   $\stackrel{h}{\rightarrow}$ &    \\
  &    $\rightarrow H^1(X,V^{\vee} \otimes L) $   &  $\rightarrow H^1(X, V \otimes V^{\vee})$ &   $\rightarrow H^1(X,V^{\vee} \otimes L^{\vee})$   $\stackrel{g}{\rightarrow}$ &  \\
  &    $\rightarrow H^2(X,V^{\vee} \otimes L) $   &  $\rightarrow H^2(X, V \otimes V^{\vee}) $ &  $\rightarrow H^2(X,V^{\vee} \otimes L^{\vee})$   $   \rightarrow  $   & \\
  &  $\rightarrow H^3(X,V^{\vee} \otimes L)   $   &  $\rightarrow H^3(X, V \otimes V^{\vee}) $ &  $\rightarrow H^3(X,V^{\vee} \otimes L^{\vee})$   $\rightarrow$ & 0.
\end{tabular}
\vspace{5pt}\\

Stability of $V$ implies $h^0(X, V \otimes V^{\vee}) =1$. Serre duality and the fact that $V =V^{\vee}$ yield, together with (\ref{h3_1}), that $h^0(X,V^{\vee} \otimes L^{\vee}) = h^3(X,V^{\vee} \otimes L) =1$. Matching the dimensions of the cohomology groups of the first line thus shows that the coboundary map $h$ has to be trivial.
Finally the dimension of the moduli space of the extension bundle is given by
\bea
h^1(X,V \otimes V^{\vee}) =  (h^1(X,L^2) -1) +  h^2(X,L^2) - {\rm rank}\, g  = h^2(X,V \otimes V^{\vee}),
\eea
where the coboundary map $g$ is given by the cup product with $H^1(X,L^2)$,
\bea
\label{map_g}
H^1(X, V \otimes L^{\vee}) \times H^1(X,L^2) \longrightarrow H^2(X, V \otimes L).
\eea
Clearly, strictly chiral recombinations with $h^1(X,L^2) = 1$ and  $h^2(X,L^2) =0$ result in bundles with no deformation moduli, while for all other cases moduli can in principle remain.
The number of remaining moduli involves in particular the rank of the map $g$, which depends on the details of the line bundle in question.
There are  certainly situations conceivable where $g$ is not of maximal rank so that unlifted moduli remain. For the minimal vector-like case where $h^1(X, L^2) = 1 = h^2(X, L^2)$ this  happens e.g. whenever the localisation inside $X$ of the various cohomology groups  appearing in (\ref{map_g}) does not allow for a non-trivial map of this type. We leave the discussion of concrete examples for future work.

\section{Discussion}

The superpotential of four-dimensional Type II orientifold
compactifications can receive non-perturbative corrections not only
from D-brane instantons invariant under the orientifold action
everywhere in moduli space, but also from objects that can become
$U(1)$ instantons for certain values of the closed string moduli. In
this article we have extended our previous analysis
\cite{Blumenhagen:2007bn} of the simplest possible type of $U(1)$
instantons with chiral intersection with the orientifold  to
multi-instanton processes involving in addition a certain type of
$O(1)$ instantons. We have shown that the specific multi-instanton
configuration can yield superpotential contributions on top of its
line of marginal stability. On the two different sides of this
hypersurface in moduli space, BPS (multi-)bound states of
different topology, but of the same total charge can form. Their
contribution to the superpotential guarantees its holomorphicity, as
in the case of $U(1)$ instantons with non-chiral intersection
analysed in \cite{GarciaEtxebarria:2007zv}. The additional
instantons in this multi-instanton configuration that allow for the
formation of BPS bound states are precisely of the type required for
lifting all extra zero modes on the line of marginal stability and
leading to a non-zero bosonic integral, and vice versa. To put
tables round, this demonstrates explicitly how the possible decay of a BPS instanton into a
non-BPS one somewhere in moduli space is encoded in its microscopic
description at marginal stability in a consistent way to prevent a
contribution to the superpotential. We have started with a pure $U(1)$ instanton at a line of marginal stability. To lift extra zero modes
we have to add new instantons. But just when the resulting multi-instanton is ready to contribute to the superpotential, the line of marginal stability
has turned into a line of threshold stability, and holomorphicity of the superpotential is ensured.

Another conclusion of our analysis is that even the class of relevant BPS objects is larger than mostly considered. The multi-instanton setup with $U(1)$ instantons maps to
Type I $D5$-brane instantons carrying holomorphic bundles and their bound states with $D1$-instantons. These are in turn S-dual to bound states of magnetised heterotic NS5-brane instantons and worldsheet instantons. Both what we called chiral and vector-like setups involving these objects have to be analysed to compute the full superpotential.
We outlined how the presence of the required couplings in the instanton worldvolume action can be determined with the help of standard algebraic techniques.
It will be interesting to check in concrete compactifications if these hitherto neglected instantons yield corrections to reckon with. This is an important question not only in view of the destabilising effect that instantons may have on four-dimensional string vacua.

\vskip 1cm
 {\noindent  {\Large \bf Acknowledgements}}
 \vskip 0.5cm
We thank O. Aharony, R. Blumenhagen, T. Brelidze, F. Denef, R. Donagi, M. Douglas, D. Joyce, T. Pantev, M. Schulz, S. Sethi and  D. Van den Bleeken for discussions and correspondence.
This research was supported in part by the National Science Foundation under
Grant No. PHY99-07949, the
Department of Energy Grant
DOE-EY-76-02-3071 and the Fay R. and Eugene L. Langberg
Endowed Chair.

\clearpage

\appendix
\section{Absence of perturbative lifting of chiral excess modes}
\label{app_Lifting}

In this appendix we further substantiate the absence of non-perturbative couplings that would lift the chiral excess modes $\lambda$ in the context of $U(1)$ instantons with chiral intersections with their orientifold image.

Consider the system of section \ref{sec_2-inst_TypeI} with intersection numbers $h^1(X,L^2)=1,$  $h^2(X,L^2)=0$.
First  we investigate if couplings of the type $ \langle \ov m \lambda^i \lambda^j  \rangle$ can exist on the locus $J \in {\cal M}_0$. For this purpose we recall that
\bea
 \ov m \in H^2(X, (L^{\vee})^2), \quad \lambda^i \in H^1(X, W \otimes L), \quad \lambda^j \in H^1(X, W^{\vee} \otimes L)
\eea
with $i=1,2$ and $j=3,4$. The above Yukawa coupling would correspond to a map
\bea
H^2(X, (L^{\vee})^2) \times H^1(X, W \otimes L) \times H^1(X, W^{\vee} \otimes L) \rightarrow {\mathbb C}.
\eea
While allowed by gauge invariance, such a map can obviously not exist in view of the degrees of the cohomology groups. This is the analogue of the worldsheet argument discussed in this context in \cite{Blumenhagen:2007bn} and reviewed in section \ref{sec_def}.
Rather we would need
\bea
H^1(X, (L^{\vee})^2) \times H^1(X, W \otimes L) \times H^1(X, W^{\vee} \otimes L) \rightarrow H^3(X, \cal O) \equiv {\mathbb C},
\eea
but by assumption $h^1(X, (L^{\vee})^2)=0$.

The absence of couplings $ \langle \ov m \lambda^i \lambda^j  \rangle$ at ${\cal M}_0$ is equivalent to the statement that the $\lambda$ modes survive as vector-like modes between the bound state $V$ and the brane $W$ as we enter into
${\cal M}_+$.
The relevant cohomology groups $H^*(X, W \otimes V)$ follow from the short exact sequence
\bea
\label{ExtWxL}
0 \rightarrow W \otimes L \rightarrow W \otimes V \rightarrow W \otimes L^{\vee} \rightarrow 0
\eea
obtained by tensoring (\ref{ExtV}) with the bundle $W$.
It
induces the long exact sequence in cohomology
\bea
0 &\rightarrow H^0(X,W \otimes L) \rightarrow H^0(X,W \otimes V) \rightarrow H^0(X,W \otimes L^{\vee}) \rightarrow \\
  &\rightarrow H^1(X,W \otimes L) \rightarrow H^1(X,W \otimes V) \rightarrow H^1(X,W \otimes L^{\vee}) \rightarrow \\
  &\rightarrow H^2(X,W \otimes L) \rightarrow H^2(X,W \otimes V) \rightarrow H^2(X,W \otimes L^{\vee}) \rightarrow \\
  &\rightarrow H^3(X,W \otimes L) \rightarrow H^3(X,W \otimes V) \rightarrow H^3(X,W \otimes L^{\vee}) \rightarrow 0.
\eea

Recall that for simplicity we take $L$ to be a line bundle.
By assumption, $W$ is stable and of zero slope for $J \in {\cal M}_0$ and also for small deformations of $J$ into ${\cal M}_+$.
Consequently,  $H^i(X,W \otimes L) = H^i(X,W \otimes L^{\vee})=0$ for $i=0,3$.
Thus the first and third lines of the long exact sequence are trivial and the sequence reduces to
\bea
0 &\rightarrow H^1(X,W \otimes L) \equiv {\mathbb C}^{n+2} \rightarrow H^1(X,W \otimes V) \rightarrow H^1(X,W \otimes L^{\vee}) \equiv {\mathbb C}^{n} \stackrel{f}{\rightarrow} \nonumber \\
  &\rightarrow H^2(X,W \otimes L)\equiv {\mathbb C}^{n} \rightarrow H^2(X,W \otimes V) \rightarrow H^2(X,W \otimes L^{\vee}) \equiv {\mathbb C}^{n+2} \rightarrow 0. \nonumber
\eea
Here we used equ. (\ref{H(WxL)}) and Serre duality for $W^{\vee} \otimes L$. The minimal zero mode situation corresponds to $n=0$.
It follows that
\bea
h^1(X,W \otimes V) &=& (n+2) + n - rk(f), \\
h^2(X,W \otimes V) &=& (n+2) + n - rk(f),
\eea
where the map $f: H^1(X, W \otimes L^{\vee}) \longrightarrow H^2(X, W \otimes L)$ is given by multiplication with the group $H^1(X,L^2)$. It can therefore be, in principle, of any rank up to $n$, depending on the concrete bundles.
In any case we see that there always exist vector-like modes counted by $H^1(X,W \otimes V)$, $H^2(X,W \otimes V)$, only some of which (namely the modes inherited from $H^2(X,W \otimes L)$ and $H^2(X,W^{\vee} \otimes L)$, which are of the wrong $U(1)_E$ charge compared to the needed excess modes in $H^1(X,W \otimes L)$ and $H^1(X,W^{\vee} \otimes L)$) can be lifted by the extension provided the map $f$ is non-zero.
In particular this means we can never lift the excess modes counted by $H^1(X,W \otimes L)$ and $H^1(X,W^{\vee} \otimes L)$.

One might wonder if there can exist couplings involving these vector-like zero modes given by $\lambda^i \in H^1(X,W \otimes V)$ and $\lambda^j \in H^1(X,W^{\vee} \otimes V)$. E.g. if W is a vector bundle with moduli $h \in H^1(X, W \otimes W^{\vee})$, one could consider couplings of the type $\langle h \lambda \lambda \rangle$, corresponding to a map
\bea
H^1(X,W \otimes V) \times H^1(X, W \otimes W^{\vee})  \times H^1(X, W^{\vee} \otimes V) \rightarrow H^3(X, \cal O) \equiv {\mathbb C}.
\eea
As $V \simeq V^{\vee}$ this map might exist. However, as discussed above,  $H^1(X,W \otimes V)$ and $H^1(X, W^{\vee} \otimes V)$ inherit their elements from  $H^1(X,W \otimes L)$ and $H^1(X, W^{\vee} \otimes L)$, and no map
\bea
H^1(X,W \otimes L) \times H^1(X, W \otimes W^{\vee})  \times H^1(X, W^{\vee} \otimes L) \rightarrow H^3(X, \cal O) \equiv {\mathbb C}
\eea
can exist.

\section{Local multi-instanton setup on $T^6/{\mathbb Z}_2\times {\mathbb Z}'_2$}
\label{App_IIA}

In this appendix we present a local realization of the multi-instanton effect discussed in the section \ref{MultiIIA} in a Type
IIA compactification. As compactification manifold we choose $
T^6/\Z_2\times \Z_2'$ orientifold with Hodge numbers
$(h_{11},h_{12})=(3,51)$ \cite{Dudas:2005jx,Blumenhagen:2005tn} which is known to
exhibit rigid cycles\footnote{For a different orientifold background
based on shift orbifolds giving rise to rigid cycles see
\cite{Blumenhagen:2006ab}.}. We adopt the notation of
\cite{Blumenhagen:2005tn}, where further details
 can be found. The orbifold
group is generated by $\theta$ and $\theta'$ acting as reflection in
the first and last two tori, respectively.

Each sector, $\theta$, $\theta'$ and $\theta\theta'$ exhibits $16$
fixed points which after blowing up give rise to additional
two-cycles with the topology of ${\mathbb P}_1$. Apart from the
usual non-rigid bulk cycles
\bea \Pi_a^B = 4 \, \bigotimes_{I=1}^3
\,(n_a^I [a^I] + \widetilde m_a^I [b^I]), \eea
defined in terms of
the fundamental one-cycles $[a^I], [b^I]$ of the $I$-th $T^2$ and
the corresponding wrapping numbers $n_a^I$ and  $\widetilde m_a^I=
m_a^I + \beta^I n_a^I$ where $\beta^I=0, 1/2$ for rectangular and
tilted tori, respectively, the background also contains
so-called $g$-twisted cycles
\bea \Pi^g_{ij} = [\alpha^g_{ij}]
\times [(n^{I_g},\widetilde{m}^{I_g})].
\eea Here ${i,j} \in
\{1,2,3,4\} \times \{1,2,3,4\}$ labels one of the 16 blown-up fixed
points of the orbifold element $g  = \theta, \theta', \theta \theta'
\in \Z_2\times \Z_2'$. These cycles are basically twice the product
of the two cycles of the corresponding ${\mathbb P}_1$ and the
$I_{g}$ invariant one cycle $[(n^{I_g},\widetilde{m}^{I_g})]$, where
$I_g = 3,1,2$ for $g  = \theta, \theta', \theta \theta'$.

Cycles which are charged under all three twisted sectors are rigid
and take the form \bea \Pi^F = \frac{1}{4} \Pi^B + \frac{1}{4}
\Bigl( \sum_{i,j \in S_{\theta}} \epsilon^{\theta}_{ij}
\Pi^{\theta}_{ij} \Bigr)+ \frac{1}{4} \Bigl(  \sum_{j,k \in
S_{\theta'}} \epsilon^{\theta'}_{jk} \Pi^{\theta'}_{jk}  \Bigr) +
\frac{1}{4} \Bigl(  \sum_{i,k \in S_{\theta \theta'}}
\epsilon^{\theta \theta'}_{ik} \Pi^{\theta \theta'}_{ik}  \Bigr).
\eea Here $S_g$ denotes the set of fixed points that the rigid brane
runs through in the $g$-twisted sector. The $\epsilon^g_{ij}=\pm 1$
correspond to the two different orientation the brane can wrap the
${\mathbb P}_1$ and have to satisfy various consistency conditions
\cite{Blumenhagen:2005tn}.

The orientifold action $\Omega\mathcal{R}$ on untwisted cycles takes
the usual form
\begin{align}
\Omega\mathcal{R}:
[(n_1,\widetilde{m}_1)(n_2,\widetilde{m}_2)(n_3,\widetilde{m}_3)]\rightarrow
[(n_1,-\widetilde{m}_1)(n_2,-\widetilde{m}_2)(n_3,-\widetilde{m}_3)]
\end{align}
whereas the twisted cycles transform as
\begin{align}
\label{Omegatwisted} \Omega\mathcal{R}:
\,\alpha^g_{ij}[(n^{I_g},\widetilde{m}^{I_g})]\rightarrow-
\eta_{\Omega\mathcal{R}}\,\eta_{\Omega\mathcal{R}g} \,
\alpha^g_{\mathcal{R}(i)\mathcal{R}(j)}[(-n^{I_g},\widetilde{m}^{I_g})],
\end{align}
where the reflection $\mathcal{R}$ leaves all fixed points of an
untilted two-torus invariant and acts on the fixed points in a
tilted two-torus as
\begin{align}
\mathcal{R}(1)=1, \qquad \mathcal{R}(2)=2, \qquad\mathcal{R}(3)=4,
\qquad\mathcal{R}(4)=3.
\end{align}
The orientifold charges $\eta_{\Omega\mathcal{R}g}=\pm 1$ are
subject to the constraint
\begin{align}
\label{etaconst}
\eta_{\Omega\mathcal{R}}\,\eta_{\Omega\mathcal{R}\theta}\,
\eta_{\Omega\mathcal{R}\theta'}\,\eta_{\Omega\mathcal{R}\theta\theta'}=-1.
\end{align}
In our subsequent local setup we choose them to be
\begin{align}
\label{eta-choice} -\eta_{\Omega\mathcal{R}}=
\eta_{\Omega\mathcal{R}\theta}
=\eta_{\Omega\mathcal{R}\theta\theta'}=
\eta_{\Omega\mathcal{R}\theta'}=1.
\end{align}
In addition we assume all three tori to be tilted such that the
orientifold planes are given by
\begin{align*}
\Pi_{O6}=-[(2,\tilde{0})(2,\tilde{0})(2,\tilde{0})]-
[(2,\tilde{0})(0,\tilde{1})(0,\tilde{1})]-2
[(0,\tilde{1})(2,\tilde{0})(0,\tilde{1})]-[(0,\tilde{1})(0,\tilde{1})(2,\tilde{0})].
\end{align*}

The $U(1)$-instanton $E$ wraps a bulk cycle of the form
\begin{align}
\Pi^B_{\Xi}=
[(-1,0)(-1,0)(-1,0)]=[(-1,-\tilde{\frac{1}{2}})(-1,-\tilde{\frac{1}{2}})(-1,-\tilde{\frac{1}{2}})]
\label{solution E}
\end{align}
and passes through the origin in all three tori. Thus its whole
homology class $\Xi$ is given by
\bea
&& \Pi^F_{\Xi}=\frac{1}{4}\Pi^B_{\Xi}+\frac{1}{4}\sum_{i,j\epsilon(13)\times(13)}
\varepsilon^{\theta}_{ij} \Pi^{\theta}_{\Xi }+
\frac{1}{4}\sum_{j,k\epsilon(13)\times(13)}
\varepsilon^{\theta'}_{jk} \Pi^{\theta'}_{\Xi }+
\frac{1}{4}\sum_{i,k\epsilon(13)\times(13)} \varepsilon^{\theta
\theta'}_{ik} \Pi^{\theta \theta'}_{\Xi }, \nonumber  \\
&& \Pi^{\theta}_{\Xi } =[((-1,-\tilde{\frac{1}{2}})], \qquad
 \Pi^{\theta'}_{\Xi } =[(-1,-\tilde{\frac{1}{2}})], \qquad \Pi^{\theta
\theta'}_{\Xi } =[(-1,-\tilde{\frac{1}{2}})].
\eea
Its orientifold image takes the form
\bea
 && \Pi^F_{\Xi'}=\frac{1}{4}\Pi^B_{\Xi}+\frac{1}{4}\sum_{i,j\epsilon(14)\times(14)}
\varepsilon^{\theta}_{ij} \Pi^{\theta}_{\Xi'}+
\frac{1}{4}\sum_{j,k\epsilon(14)\times(14)}
\varepsilon^{\theta'}_{jk} \Pi^{\theta'}_{\Xi'}+
\frac{1}{4}\sum_{i,k\epsilon(14)\times(14)} \varepsilon^{\theta
\theta'}_{ik} \Pi^{\theta \theta'}_{\Xi'} , \nonumber \\
&&\Pi^B_{\Xi'}=[(-1,\tilde{\frac{1}{2}})(-1,\tilde{\frac{1}{2}})(-1,\tilde{\frac{1}{2}})],
\\ \nonumber
&& \Pi^{\theta}_{\Xi'}=[((-1,\tilde{\frac{1}{2}})], \qquad
\Pi^{\theta'}_{\Xi'}=[(-1,\tilde{\frac{1}{2}})], \qquad \Pi^{\theta
\theta'}_{\Xi'}=[(-1,\tilde{\frac{1}{2}})]. \eea With the
intersection formulae \bea \nonumber && \Pi^B_a \circ \Pi^B_b= 4
\prod^3_{i=1} (n^i_a
\widetilde{m}^i_b-n^i_b \widetilde{m}^i_a), \\
&&\Pi^g_{ij} \circ \Pi^h_{kl}= 4 \delta^{gh} \, \delta^{ik}\,
\delta^{jl} (n^{I_g}_a \widetilde{m}^{I_h}_b-n^{I_h}_b
\widetilde{m}^{I_g}_a) \eea it is easy to show that \bea
\Pi_{\Xi'}\circ \Pi_{\Xi}= 1, \qquad \Pi_{O6}\circ  \Pi_{\Xi}=1.
\eea As discussed in section \ref{sec_def} we need $4$ additional
charged zero modes between $E$ and some D-brane $a$. We choose the
brane $a$ wrapping the cycle \bea
\Pi^F_a=\frac{1}{4}\Pi^B_{a}+\frac{1}{4}\sum_{i,j\epsilon(12)\times(34)}
\varepsilon^{\theta}_{ij} \Pi^{\theta}_a+
\frac{1}{4}\sum_{j,k\epsilon(12)\times(34)}
\varepsilon^{\theta'}_{jk} \Pi^{\theta'}_a+
\frac{1}{4}\sum_{i,k\epsilon(34)\times(34)} \varepsilon^{\theta
\theta'}_{ik} \Pi^{\theta \theta'}_a \label{E} \eea with \bea
&& \Pi^B_{a}=[(0,1)(-4,3)(-4,3)]=[(0,\tilde{1})(-4,\tilde{1})(-4,\tilde{1})],\,\,\,\\
&& \Pi^{\theta}_a=[(-4,\tilde{1})], \qquad
\Pi^{\theta'}_a=[(0,\tilde{1})], \qquad \Pi^{\theta
\theta'}_a=[(-4,\tilde{1})]. \eea Note that in contrast to section
\ref{sec_NPlifting1} we choose the D-brane to be not invariant under
the orientifold action. Thus we additionally have to ensure
$\Pi_{\Xi}\circ \Pi_a'=0$ for its orientifold image $a'$, which is
indeed satisfied.  In order to satisfy supersymmetry we choose the
complex structure moduli $U^I$ to be
\begin{align}
U^{1}= \frac{8}{3}\,\,, \qquad \qquad U^{2}=4\,\,, \qquad \qquad
U^{3}=4\,\,. \label{complex structure moduli}
\end{align}

As described in \cite{Blumenhagen:2007bn} the ${\overline
\tau}_{\dot{\alpha}}$- and ${\overline \mu}_{\dot{\alpha}}$-modes
can be soaked up by the coupling $m  {\overline
\tau}_{\dot{\alpha}}{\overline \mu}^{\dot{\alpha}} $ but there is no
way to absorb the charged zero modes $\lambda$ unless we take into
account additional $O(1)$-instantons. Indeed there are two
instantons satisfying the constraints \eqref{Int1}. Their homology
classes are given by
\begin{align}
\label{E1}
&&\Pi^F_{\widetilde{\Xi}_1}=\frac{1}{4}\Pi^B_{\widetilde{\Xi}_1}+\frac{1}{4}\sum_{i,j\epsilon(12)\times(12)}
\varepsilon^{\theta}_{ij} \Pi^{\theta}_{\widetilde{\Xi}_1}+
\frac{1}{4}\sum_{j,k\epsilon(12)\times(12)}
\varepsilon^{\theta'}_{jk} \Pi^{\theta'}_{\widetilde{\Xi}_1}+
\frac{1}{4}\sum_{i,k\epsilon(12)\times(12)} \varepsilon^{\theta
\theta'}_{ik} \Pi^{\theta\theta'}_{\widetilde{\Xi}_1}, \\
&&\Pi^F_{\widetilde{\Xi}_2}=\frac{1}{4}\Pi^B_{\widetilde{\Xi}_2}+\frac{1}{4}\sum_{i,j\epsilon(34)\times(12)}
\varepsilon^{\theta}_{ij} \Pi^{\theta}_{\widetilde{\Xi}_2}+
\frac{1}{4}\sum_{j,k\epsilon(12)\times(12)}
\varepsilon^{\theta'}_{jk} \Pi^{\theta'}_{\widetilde{\Xi}_2}+
\frac{1}{4}\sum_{i,k\epsilon(34)\times(12)} \varepsilon^{\theta
\theta'}_{ik} \Pi^{\theta\theta'}_{\widetilde{\Xi}_2}
\end{align}
with
\begin{align}
\Pi^B_{\widetilde{\Xi}_1}=\Pi^B_{\widetilde{\Xi}_2}=
[(2,-1)(2,-1)(2,-1)]=[(2,\tilde{0})(2,\tilde{0})(2,\tilde{0})], \qquad \,\,\,  \\
\Pi^{\theta}_{\widetilde{\Xi}_1}=\Pi^{\theta}_{\widetilde{\Xi}_2}=
[(2,\tilde{0})], \qquad
\Pi^{\theta'}_{\widetilde{\Xi}_1}=\Pi^{\theta'}_{\widetilde{\Xi}_2}=
[(2,\tilde{0})], \qquad
\Pi^{\theta\theta'}_{\widetilde{\Xi}_1}=\Pi^{\theta\theta'}_{\widetilde{\Xi}_2}=
[(2,\tilde{0})].
\end{align}
Note that both cycles are invariant under the orientifold action and
are separated in the first torus ensuring that the additional zero
modes appearing in the $\widetilde{E}_1 - \widetilde{E}_2$ sector
become massive. Now it is possible to soak up all the zero modes via
the couplings \eqref{S_1}, \eqref{S_2} and \eqref{S_3}.

Let us briefly discuss the holomorphicity of the superpotential
based on this example. The Yukawa couplings $\ov L_i$ and $Y_i$ in
\eqref{S_1} and  \eqref{S_2}, respectively take the form \bea Y_{k
ij}=y_{kij} \,\,\prod^3_{I=1}
\Gamma^{\frac{1}{4}}_{1+\phi^I_{Ea},1-\phi^I_{E\tilde{E}_k},
1-\phi^I_{\tilde{E}_k a }}, \quad \overline{L}_k=l_k \,\prod^3_{I=1}
\Gamma^{\frac{1}{4}}_{-\phi^I_{EE'},\phi^I_{E\tilde{E}_k},\phi^I_{E\tilde{E}_k}},
\label{nh yukawa} \eea where $\phi^I_{ij}$ denotes the intersection
angle between instanton $i$ and brane or instanton $j$, respectively
and
\begin{align}
 \Gamma_{\alpha,\,\beta,\,\gamma}=
 \frac{\Gamma(1-\alpha)\,\Gamma(1-\beta)\,
\Gamma(1-\gamma)}{\Gamma(\alpha)\Gamma(\beta) \Gamma(\gamma)}.
\end{align}
The lowercase letters denote the holomorphic part of the Yukawa
couplings, which essentially are given by the world sheet
contributions. Note that the dependence on $\phi_{E\tilde{E}_k}$ in
\eqref{bos1} and \eqref{bos2} drops out due to the inverse
dependence of $Y^2_{ikl}$ to $\overline{L}_i$. In addition there are
also non-holomorphic contributions from the annulus diagrams
$A(E,a)$ and $A(\tilde{E}_k,a)$ as well as the M\"obius diagram
$M(E,O6)$ \cite{Blumenhagen:2006xt}. In our example they are given
by \cite{Akerblom:2007uc, Akerblom:2007np, Blumenhagen:2007ip}
\bea\nonumber
\exp(A^{n.h.}(E,a))&= \left( \frac{\Gamma(1+\phi^1_{Ea})
\Gamma(1+\phi^2_{Ea}) \Gamma(1+\phi^3_{Ea})}{\Gamma(-\phi^1_{Ea})
\Gamma(-\phi^2_{Ea}) \Gamma(-\phi^3_{Ea})}\right)
,\\ \label{nh loop}
\exp(A^{n.h.}(\tilde{E}_{k},a))&= \left(
\frac{\Gamma(\phi^1_{\tilde{E}_{k}a})
\Gamma(\phi^2_{\tilde{E}_{k}a})
\Gamma(\phi^3_{\tilde{E}_{k}a})}{\Gamma(1-\phi^1_{\tilde{E}_{k}a})
\Gamma(1-\phi^2_{\tilde{E}_{k}a})
\Gamma(1-\phi^3_{\tilde{E}_{k}a})}\right)^{\frac{1}{2}},\\ \nonumber
\exp(M^{n.h.}(E,O6))&=\left(
 \frac{\Gamma(-\phi^1_{E'E}) \Gamma(-\phi^2_{E'E})
\Gamma(-\phi^3_{E'E})}{\Gamma(1+\phi^1_{EE'}) \Gamma(1+\phi^2_{EE'})
\Gamma(1+\phi^3_{EE'} }\right)^{\frac{1}{2}}.
\eea
Indeed, after plugging \eqref{nh yukawa} and \eqref{nh
loop} into \eqref{bos1} or \eqref{bos2} all angle dependence cancels and one is left with a holomorphic expression for the superpotential.

Let us deform the complex structure in the first torus away from the
line of marginal stability. Note that under deformation of the
complex structure $U^1$, while keeping the complex structures in the
other two tori fixed, the brane $a$ remains supersymmetric. For
$U^1>8/3 $ we induce a positive Fayet-Iliopoulus parameter $\xi$ for
the $U(1)_{E}$ and as described in section \ref{fayet} the cycles
$\Xi$ and $\Xi'$ combine into a new special Lagrangian $Y=\Xi
\#\Xi'$ preserving the same ${\cal N}=1$ supersymmetry as the
orientifold. The whole multi-instanton configuration is then given
by $(\Xi \#\Xi') \cup \widetilde{\Xi}_1 \cup \widetilde{\Xi}_2$. For
$U^1<8/3$ we induce a negative $\xi$ for the $U(1)_E$ and the
multi-instanton configuration recombines into the new BPS state
$\widetilde{\Xi}_1 \# ((\Xi \cup \Xi')) \# \widetilde{\Xi}_2$.

\clearpage
\nocite{*}
\bibliography{rev}
\bibliographystyle{utphys}

\end{document}